\documentclass[10pt,aps,prc,floatfix,twocolumn,superscriptaddress,nofootinbib]{revtex4-2}

\usepackage[table,dvipsnames]{xcolor}
\usepackage{amsfonts,amsmath,amssymb,bm}
\usepackage{graphicx}
\usepackage[utf8]{inputenc}
\usepackage{xspace}
\usepackage{dcolumn}
\usepackage{braket}
\usepackage{multirow}
\newcolumntype{d}[1]{D{.}{.}{#1}}
\usepackage{cellspace}
\usepackage{isotope}
\usepackage{xparse}
\usepackage{physics}
\usepackage{dsfont}
\usepackage[pdfpagelabels, pdfencoding=auto, psdextra]{hyperref}
\hypersetup{%
 pdftitle={Active learning emulators for nuclear two-body scattering in momentum space},
 pdfsubject={Active learning emulators for nuclear two-body scattering in momentum space},
 pdfkeywords={low-energy nuclear theory, machine learning, emulators, surrogate models, reduced basis method, nuclear scattering, active learning, greedy algorithm, DOE STREAMLINE collaboration}, 
 unicode = true,
 breaklinks = true,
 colorlinks = true,
 linkcolor = blue,
 menucolor = blue,
 citecolor = blue,
 urlcolor = blue
}
\usepackage{hyperxmp}
\usepackage{enumitem}
\usepackage{mathtools}
\usepackage{bm}

\graphicspath{{./figures/}}

\newcommand {\nc} {\newcommand}
\nc {\IR} [1]{\textcolor{red}{#1}}
\nc {\IB} [1]{\textcolor{blue}{#1}}         
\nc {\IP} [1]{\textcolor{magenta}{#1}}
\nc {\IM} [1]{\textcolor{Bittersweet}{#1}}  
\nc {\IE} [1]{\textcolor{Plum}{#1}}         
\nc {\IG} [1]{\textcolor{OliveGreen}{#1}}

\newcommand{\orcid}[1]{\href{https://orcid.org/#1}{\includegraphics[scale=0.055]{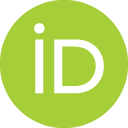}}}

\newcommand{\paramVec}{\vb*{\theta}} 

\newcommand{\ntheta}{n_{\paramVec}}
\newcommand{\Amat}{\vb{A}}  
\newcommand{\tAmat}{\widetilde\Amat}
\newcommand{\bvec}{\vb{b}}   
\newcommand{\cvec}{\vb{c}}   

\newcommand{\tbvec}{\widetilde\bvec}

\newcommand{\Xmat}{\vb{X}}  
\newcommand{\Ymat}{\vb{Y}}  

\newcommand{\residual}{\vb{r}}

\newcommand{\exactErrorVec}{\vb{e}} 
\newcommand{\romResidual}{\vb{r}}
\newcommand{\romResidualY}{\rvec_{\vb{Y}}}
\newcommand{\rvec}{\vb{r}}   

\newcommand{\jax}{\texttt{JAX}\xspace}

\newcommand{\oneSzero}{${}^1\text{S}_0$\xspace}  %
\newcommand{\threeSone}{${}^3\text{S}_1$\xspace}  %
\newcommand{\threeDone}{${}^3\text{D}_1$\xspace}  %
\newcommand{\threeSDone}{\threeSone--\threeDone}  %

\newcommand{\fmi}{\, \text{fm}^{-1}}
\newcommand{\fm}{\, \text{fm}}
\newcommand{\MeV}{\, \text{MeV}}

\newcommand{\nTwoLO}{N$^2$LO\xspace}

\newcommand{\pr}{p}
\newcommand{\given}{\, | \,}
\newcommand{\tol}{\alpha}

\newcommand{\rev}[1]{#1}

\begin{document}

\title{Active learning emulators for nuclear two-body scattering in momentum space}

\author{A.~Giri~\orcid{0009-0000-7239-1782}}
\email{ag086822@ohio.edu}
\affiliation{Department of Physics and Astronomy, \href{https://ror.org/01jr3y717}{Ohio University}, Athens, Ohio~45701, USA}

\author{J.~Kim~\orcid{0009-0003-9488-0247}}
\email{jane.kim@anl.gov}
\affiliation{Department of Physics and Astronomy, \href{https://ror.org/01jr3y717}{Ohio University}, Athens, Ohio~45701, USA}
\affiliation{Physics Division, \href{https://ror.org/05gvnxz63}{Argonne National Laboratory}, Lemont, Illinois~60439, USA}

\author{C.~Drischler~\orcid{0000-0003-1534-6285}}
\email{drischler@ohio.edu}
\affiliation{Department of Physics and Astronomy, \href{https://ror.org/01jr3y717}{Ohio University}, Athens, Ohio~45701, USA}
\affiliation{\href{https://ror.org/03r4g9w46}{Facility for Rare Isotope Beams}, \href{https://ror.org/05hs6h993}{Michigan State University}, East Lansing, Michigan~48824, USA}

\author{Ch.~Elster~\orcid{0000-0002-2459-1226}}
\email{elster@ohio.edu}
\affiliation{Department of Physics and Astronomy, \href{https://ror.org/01jr3y717}{Ohio University}, Athens, Ohio~45701, USA}

\author{R.~J.~Furnstahl~\orcid{0000-0002-3483-333X}}
\email{furnstahl.1@osu.edu}
\affiliation{Department of Physics, \href{https://ror.org/00rs6vg23}{The Ohio State University}, Columbus, Ohio~43210, USA}

\date{\today}

\begin{abstract}  

We extend the active learning emulators for two-body scattering in coordinate space with error estimation, recently developed by Maldonado~et~al.\ [\href{https://journals.aps.org/prc/abstract/10.1103/k77q-f82l}{Phys.~Rev.~C \textbf{112}, 024002}], to coupled-channel scattering in momentum space. Our full-order model (FOM) solver is based on the Lippmann-Schwinger integral equation for the scattering $t$-matrix as opposed to the radial Schr\"odinger equation. We use (Petrov-)Galerkin projections and high-fidelity calculations at a few snapshots across the parameter space of the interaction to construct efficient reduced-order models (ROMs), trained by a greedy algorithm for locally optimal snapshot selection. Both the FOM solver and the corresponding ROMs are implemented efficiently in Python using Google's \texttt{JAX} library. We present results for emulating scattering phase shifts in coupled and uncoupled channels and cross sections, and assess the accuracy of the developed ROMs and their computational speedup factors. We also develop emulator error estimation for both the $t$-matrix and the total cross section. The software framework for reproducing and extending our results is publicly available. Together with our recent advances in developing active learning emulators for three-body scattering, these emulator frameworks set the stage for full Bayesian calibrations of chiral nuclear interactions and optical models against scattering data with quantified emulator errors.

\end{abstract}

\maketitle

\section{Introduction} 

A central challenge in modern nuclear theory is the rigorous quantification of uncertainty, typically achieved through Bayesian statistical methods. 
These methods require repeated evaluations of high-fidelity models, which can be computationally expensive or even prohibitively slow depending on the application.
To overcome this challenge, fast and accurate emulators have recently gained significant popularity in the nuclear physics community.
These surrogate models can provide approximate yet highly accurate reduced-order representations of the corresponding high-fidelity (i.e., full-order) models at significantly reduced computational cost. 
Comprehensive review articles on emulators and their applications to nuclear structure and scattering calculations can be found in Refs.~\cite{Melendez:2022kid,Drischler:2022ipa,Duguet:2023wuh}.\footnote{\rev{For original work on projection-based emulators for bound systems, we refer the reader to, e.g., Refs.~\cite{Frame:2017fah,Ekstrom:2019lss,Konig:2019adq,Demol:2019yjt}. 
A promising alternative approach to accelerating scattering calculations was developed in Ref.~\cite{Miller:2021vby} using the wave-packet continuum discretization method and GPUs.}}

Among these emulator techniques, the reduced basis method (RBM) has emerged as a powerful and flexible approach. 
The RBM constructs low-dimensional approximations of the FOM's solution space using a set of high-fidelity calculations, also called snapshots~\cite{hesthaven2015certified,Quarteroni:218966,Benner20201}.
These snapshot calculations can be placed na\"ively in the parameter space, e.g., using Latin hypercube sampling (LHS) combined with orthonormalization and compression facilitated by the proper orthogonal decomposition (POD), as illustrated  \rev{by Maldonado et al.}\ in Figure~1 of Ref.~\cite{Maldonado:2025ftg}. 
Alternatively, they can be selected using active learning, applying a so-called greedy algorithm~\cite{Sarkar:2021fpz,Bonilla:2022rph,Maldonado:2025ftg} that iteratively places snapshot calculations where the (estimated) emulator error is the largest. 
It thereby iteratively minimizes the error until a desired accuracy is achieved, preferably with fewer snapshot calculations.

In Ref.~\cite{Maldonado:2025ftg}, the latter approach was developed and tested for nucleon-nucleon (NN) scattering with chiral interactions in coordinate space, which has been the prototypical test case for emulators in general~\cite{Furnstahl:2020abp,Melendez:2021lyq,Drischler:2021qoy,Odell:2023cun,Garcia:2023slj,Catacora-Rios:2025pau}. 
While fast and accurate emulators for NN scattering are appealing in practice, such emulators are crucial for higher-body scattering, such as nucleon-deuteron (Nd) scattering~\cite{Kievsky2008,Marcucci2019,Witala:2021xqm,Gnech:2025gsy,Gnech:2025lbg}. 
Recently, \rev{Refs.~\cite{Gnech:2025gsy,Gnech:2025lbg}} demonstrated the efficacy of this active learning approach for training proton-deuteron scattering emulators below the deuteron breakup threshold energy, a crucial step toward full Bayesian calibration of chiral three-nucleon (3N) interactions.

Here, we build on the active-learning emulators for NN scattering in uncoupled partial-wave channels with error estimation, initially developed in Ref.~\cite{Maldonado:2025ftg}. 
We extend them in the following essential ways:
\begin{itemize}
    \item scattering in momentum space, in both coupled and uncoupled partial-wave channels, giving access to a wide range of modern chiral interactions;
    \item proof-of-principle estimation of emulator errors for total cross sections and application of these emulators to Bayesian calibration of chiral NN potentials;
    \item high-performance computing (HPC) implementation of these emulators and assessment of their speedups.
\end{itemize}
Together with the 3N emulators developed in Ref.~\cite{Gnech:2025gsy,Gnech:2025lbg}, this paper provides fast and accurate emulators for calibrating chiral NN and 3N forces efficiently to scattering observables.

To this end, we consider the Lippmann-Schwinger (LS) integral equation in momentum space for the scattering $t$-matrix as our FOM.
The FOM can be efficiently solved by discretizing the LS equation using an accurate quadrature rule for the principal value integration, resulting in a system of coupled linear equations for the half-on-shell and on-shell matrix elements of the $t$-matrix~\cite{Landau:1997}.
We then use Galerkin and Petrov-Galerkin projections, along with high-fidelity calculations at a few snapshots of the nuclear interaction parameter space obtained by the greedy algorithm, to construct fast and accurate scattering emulators.
Google's library \jax~\cite{jax2018github} enables the efficient implementation of both the FOM solver and the corresponding reduced-order model (ROM) in Python.
We demonstrate that Bayesian parameter estimation of nuclear potentials with quantified emulator errors is feasible using our emulators, setting the stage for their applications to rigorous calibrations of chiral nuclear forces.
We present results for emulating cross sections and scattering phase shifts in coupled and uncoupled channels using chiral NN interactions and assess the accuracy of the developed ROMs and their computational speedup factors.

The remainder of this paper is organized as follows.
In Sec.~\ref{sec:formalism}, we describe the formalism of our scattering emulators, which consist of a $t$-matrix high-fidelity solver in momentum space and corresponding ROMs obtained via (Petrov-)Galerkin projections. 
We then discuss the greedy algorithm in momentum space for the Minnesota potential with a two-dimensional parameter space, which serves as a test case. 
Next, we present in Sec.~\ref{sec:results} our main results for scattering phase shifts and cross-sections based on more realistic chiral potentials. 
We also discuss the efficient implementation of our FOM and ROM solvers in Python and discuss the ROM's speedup factors.
Finally, we perform a proof-of-principle Bayesian parameter calibration of a chiral interaction~\cite{Gezerlis:2014zia} to total scattering cross-sections, with emulator errors estimated using the developed methods.
This article concludes in Sec.~\ref{sec:summary} with an overview and an outlook.
Appendices~\ref{appendixA} and~\ref{appendixB} provide additional information on calculating phase shifts, inelasticity parameters, and total cross sections in momentum space.

We use natural units in which $\hbar = c = 1$ and denote matrices and vectors with uppercase and lowercase letters, respectively, both typeset in boldface. 
The software for reproducing and extending our results \rev{is} publicly available on GitHub~\cite{BUQEYEsoftware}.

\section{Formalism}
\label{sec:formalism}

\begin{table}[tb]
\renewcommand{\arraystretch}{1.2}
\setlength{\tabcolsep}{6pt}
\caption{
Notation used in this work.
}
\label{tab:notation}
\begin{ruledtabular}
\begin{tabular}{lp{5.7cm}} 
Notation & Description \\ 
\colrule 
$n_b$ & number of reduced basis elements \\
$\paramVec$, $n_\theta$ & contains the $\ntheta$ low-energy couplings or other model parameters as its components  \\
$a$ & summation index that runs over affine components (i.e., LECs)\\
$N$ & number of momentum grid points \\
$\Xmat$ & snapshot matrix \\
$k,k'$ & initial and final relative momenta\\
$k_0$, $E_{k_0}$, $E_\text{lab}$ & on-shell momentum, associated center-of-mass, and laboratory kinetic energies \\
$T^j_{\ell' \ell} (k, k'; E_{k_0})$ & partial-wave decomposed $T$-matrix\\
$S^j_{\ell' \ell} (k, k'; E_{k_0})$ & partial-wave decomposed $S$-matrix\\
$\tau^j_{\ell \ell'}(E_{k_0})$ & partial-wave decomposed dimensionless scattering amplitude\\
$\delta^j_{\ell \ell'}(E_{k_0})$ & scattering phase shift \\
$V^j_{\ell',\ell} (k, k')$ &  partial-wave decomposed potential\\
$\mu$ & reduced nucleon mass\\
$\cvec(\paramVec)$ & coefficient vector of the ROM\\
$j,\ell,\ell'$ & angular momentum quantum numbers of the initial and final states\\
$\pr(x)$ & probability distribution of the variable $x$\\
$\vb{t}$, $\tilde{\vb{t}}$ & FOM and ROM vector containing the discretized $t$-matrix, respectively \\
$\tol$ & requested error tolerance in the emulated $\vb{t}$-matrix when training the emulator
\end{tabular}
\end{ruledtabular}
\end{table}

In this section, we discuss the formal aspects of solving the scattering equation for the NN system, setting the stage for the high-fidelity solutions that the reduced-order models aim to emulate. 
For the reader's convenience, we will use general expressions and provide details in the appendices. 
The notation is summarized in Table~\ref{tab:notation}.

\subsection{Full-Order Model: High-Fidelity solution of the NN \texorpdfstring{$t$}{t}-matrix equation}

In its most general form, the transition amplitude for scattering is given by the Lippman-Schwinger (LS) equation in operator form,
\begin{equation} \label{eq:2.1}
\widehat{T}(z) =\widehat{V} + \widehat{V} \widehat{G}_0(z) \widehat{T}(z),
\end{equation}
where $\widehat{V}$ represents the interaction and $\widehat{G}_0(z) = (z - \widehat{H}_0)^{-1}$, with $z = E + i\varepsilon$ and $\varepsilon > 0$ defining the propagator with outgoing boundary conditions. 
The operator $\widehat{H}_0$ stands for the free Hamiltonian, which is given by $\widehat{H}_0={\widehat p}/2\mu$, with ${\widehat p}$ being the momentum operator and $\mu$ the reduced mass of the NN system.
\rev{The operator structure of any NN potential is such that while the total angular momentum $j$ is conserved, the orbital angular momentum $\ell$ is not conserved due to tensor forces.}
Thus, an angular momentum decomposition of Eq.~\eqref{eq:2.1} leads to a single integral equation if $\ell = \ell'$, 
\begin{multline} \label{eq:2.2}
T^j_\ell(k, k'; E_{k_0}) = V^j_\ell(k, k') \\
 + \lim_{\varepsilon \to 0} \int_0^\infty dk'' \, k''^2 \frac{V^j_\ell(k, k'') T^j_\ell(k'', k'; E)}{E_{k_0} - E'' + i\varepsilon},
\end{multline}
with $E_{k_0}=k_0^2/2\mu$ being the center-of-mass energy, and $k_0$ the on-shell momentum. 
The operators in the NN interaction that are of tensor character only preserve the total angular momentum $j = \ell \pm 1$, and therefore, the angular momentum decomposition leads to coupled equations,
\begin{multline} \label{eq:2.3}
 T^j_{\ell \ell'}(k,k'; E_{k_0})  =  V^j_{\ell \ell'}(k,k') \\ 
 + \sum_{\ell''} \lim_{\varepsilon \to 0} \int_0^{\infty} dk'' \, k''^2  \frac{V^j_{\ell \ell''}(k,k'') T^j_{\ell'' \ell'}(k'',k';E_{k_0})}{E_{k_0} - E'' + i\varepsilon} ,
\end{multline}
which simplifies to Eq.~\eqref{eq:2.2} without channel coupling (i.e., $\ell = \ell'$).
The solution and numerical implementation of two-body partial-wave $t$-matrices are well-known in the literature, e.g., as given in Ref.~\cite{Glockle:1983}. 
However, we describe our specific implementation in Appendix~\ref{appendixA} for completeness.

For an efficient offline-online decomposition of our emulator~\cite{Drischler:2022ipa,Duguet:2023wuh}, we require that the dependence on the parameters $\paramVec$ of the NN potential $V_{\ell,\ell'}(k,k'; \paramVec)$ is affine so that the potential's parameter dependence can be separated from its momentum dependence. 
That means that the partial-wave decomposed potential,
\begin{equation} \label{eq:affine_potential}
    V_{\ell,\ell'}(k,k'; \paramVec) = \sum_{a=0}^{\ntheta} h_a^{(\ell,\ell')}(\paramVec) \mathcal{V}_a^{(\ell,\ell')}(k,k'),
\end{equation}
can be written as sums of products of the (smooth) parameter-dependent functions $h_a^{(\ell,\ell')}(\paramVec)$ and the parameter-independent functions $\mathcal{V}_a^{(\ell,\ell')}(k,k')$.
In Eq.~\eqref{eq:affine_potential}, we introduced the auxiliary component with constant value $\paramVec_0 \equiv 1$ to accommodate parameter-independent terms in the potential~\eqref{eq:affine_potential}. 
In this work, this extra dimension will encode all pion-exchange contributions to $V_{\ell,\ell'}(k,k')$, while the other components of the parameter vector $\paramVec$ consist of all short-range low-energy constants (LECs).%
\footnote{In the case of chiral interactions, the parameter dependence on the LECs is even linear; i.e., $h_a^{(\ell,\ell')}(\paramVec) = \paramVec_a$.}
We refer to the discussion of the empirical interpolation method (EIM) in, e.g., Refs.~\cite{Odell:2023cun,Catacora-Rios:2025pau} and references therein, which can be applied to render potentials with non-affine parameter dependencies approximately affine.
Hence, we expect our emulators to cover a wide range of momentum-space potentials, including chiral interactions and optical models. 
\rev{In what follows, we focus on chiral interactions in coupled and uncoupled channels.} 

We only need the half-shell partial-wave $t$-matrices to construct the ROMs. 
Therefore, we can replace the momentum $k'$ in Eqs.~\eqref{eq:2.2} and~\eqref{eq:2.3} by the (fixed) on-shell momentum $k_0$, and the single-channel half-shell partial-wave $t$-matrix $T^j_{\ell}(k,k_0; E_{k_0})$ becomes a vector of length ($N+1$) when discretized on a momentum grid of size $N$.
Here, we choose $N=80$ to obtain high-accuracy results \rev{and truncate the numerical integration at a sufficiently high-enough maximum momentum, $k_\mathrm{max} = 20 \fmi$}.
In the coupled-channel case, the half-shell partial wave $t$-matrix for a given total angular momentum $j$, $T^j_{\ell \ell'}(k,k_0; E_{k_0})$, with $\ell, \ell'= j\mp 1$, will be a vector of length $2(N+1) \times 2(N+1)$ in its discretized form on a momentum grid of size $N$.  
However, due to the structure of the emulator it is advantageous to rewrite the coupled channel half-shell $t$-matrix as a vector of length $4(N+1)$ as $(T^j_{mm},T^j_{mp},T^j_{pm},T^j_{pp})$, where the subscripts indicate the different channels as laid out in Appendix~\ref{appendixA}. 

To obtain partial wave $S$-matrix elements, one only needs the on-shell $t$-matrix elements, $ T^j_{\ell}(k_0,k_0; E_{k_0})$ for the single channels, and $T^j_{\ell,\ell'}(k,k_0; E_{k_0})$ for the coupled channels. 
We define the partial wave $S$-matrix elements as
\begin{equation} \label{eq:2.5}
S^j_{\ell \ell'}(E_{k_0}) = 1 + 2i \tau^j_{\ell \ell'}(E_{k_0}),
\end{equation}
where 
\begin{equation}
\tau^j_{\ell \ell'}(E_{k_0}) = -\frac{\pi}{2} (2\mu) k_0 T^j_{\ell \ell'}(k_0,k_0; E_{k_0})
\label{eq:2.6}
\end{equation}
corresponds to the dimensionless scattering amplitude. 
In terms of phase shifts $\delta^j_{\ell \ell'}(E_{k_0})$, $S^j_{\ell \ell'}(E_{k_0})$ reads~\cite{Joachain} 
\begin{equation} \label{eq:2.7}
S^j_{\ell \ell'}(E_{k_0}) = \eta^j_{\ell \ell'}(E_{k_0}) e^{2i \delta^j_{\ell \ell'}(E_{k_0})},
\end{equation}
where $\delta^j_{\ell \ell'}(E_{k_0})$ is the scattering phase shift and $\eta^j_{\ell \ell'}(E_{k_0})$ the ``inelasticity'' or ``absorption'' factor that has to be one for elastic scattering with real-valued potentials, which is the case considered here. 
That is, the condition $\eta^j_{\ell \ell'}(E_{k_0})=1$ represents the unitarity of the $S$-matrix. 
We will use this condition to test if the solutions constructed by the emulator fulfill unitarity. 
For the explicit calculation of phase shifts and inelasticity parameters from the single- and coupled-channel $S$-matrix elements, we refer to Appendix~\ref{appendixB}, where we follow the Stapp convention~\cite{Stapp:1956mz} to obtain the coupled-channel eigen-phase shifts.  

To obtain the FOM, we perform a partial-wave decomposition of the $t$-matrix and write the resulting integral equation as a system of linear equations as outlined in Appendix~\ref{appendixA}. 
The matrix form of the LS equation for single channel scattering is given in Eq.~\eqref{eq:A.12} and the one for coupled channel scattering in Eq.~\eqref{eq:A.19}. Both matrix equations are of the form,
\begin{equation} \label{eq:linear_system}
    \Amat(\paramVec) \vb{t}(\paramVec) = \bvec(\paramVec) \,.
\end{equation}
Here, we explicitly emphasize the dependence of the matrix elements on the parameters $\paramVec$. 
The vector $\vb{t}$ contains the half-shell $t$-matrix elements evaluated on the quadrature grid and the on-shell $t$-matrix element. 
Once the linear system~\eqref{eq:linear_system} is solved for $\vb{t}$ using standard linear algebra methods, we obtain scattering phase shifts at center-of-mass energy $E_{k_0}$ in a given partial-wave channel using Eqs.~\eqref{eq:2.5}--\eqref{eq:2.7}.
Furthermore, to compute high-fidelity scattering cross sections, we use the optical theorem, which results in~\cite{Silbar:1979wp}
\begin{equation} \label{eq:tot_cs}
    \sigma_{\text{tot}} = -\frac{i\pi}{k_0^2} \sum_{j} (2j+1) \sum_{k}\tau^j_{k} (E_{k_0}). 
\end{equation}
Here, the sum with index $k$ runs over the singlet, triplet, and coupled channels.
The amplitudes $\tau^j_\pm (E_{k_0})$ are defined similarly to the scattering $t$-matrix in Eq.~\eqref{eq:B.1}. 
More information can be found in Appendix~\ref{appendixB}.

\subsection{Reduced-Order Models: (Petrov-)Galerkin projections}

\rev{We briefly summarize in this section how ROMs are built using Galerkin projections (G-ROM) and Least-Squares Petrov-Galerkin projections (LSPG-ROM).} 
For a given $\paramVec$, the FOM of Eq.~\eqref{eq:linear_system} can be approximated by
\begin{equation} \label{eq:rom_reduction}
    \vb{t}(\paramVec) \approx \tilde{\vb{t}}(\paramVec) = 
    \Xmat \cvec(\paramVec) \,.
\end{equation}
Here, $\Xmat$ is the snapshot matrix that has $n_b$ column vectors spanning the reduced space of the emulator. 
The columns of $\Xmat$ are orthonormalized so that $\Xmat^\dagger\Xmat$ is the identity matrix. 
The G-ROM approach finds the coefficient vector $\cvec(\paramVec)$ in Eq.~\eqref{eq:rom_reduction} by constraining the residual vector, 
\begin{equation}\label{eq:residual}
    \residual(\paramVec) = \bvec(\paramVec) - \Amat(\paramVec) \tilde{\vb{t}}(\paramVec) \,,
\end{equation}
to be orthogonal to the reduced space, which is: $\Xmat^\dagger \residual(\paramVec) = \vb{0}$. This leads to the G-ROM equations:
 \begin{multline} \label{eq:G-ROM_lin_system}
        \tAmat(\paramVec) \cvec(\paramVec) = \tbvec(\paramVec) \\ 
        \text{with} \quad 
    \tAmat(\paramVec) = \Xmat^\dagger \Amat(\paramVec) \Xmat \,, \quad 
    \tbvec(\paramVec) = \Xmat^\dagger \bvec(\paramVec) \,.
 \end{multline}
Here, $\boldsymbol{A}$ is an $(N+1) \times (N+1)$ matrix, while $\boldsymbol{\tilde{A}}$ is an $n_b \times n_b$ matrix, with $n_b  \ll  N$. Similarly, $\boldsymbol{b}$ is a vector of length-$(N+1)$, while $\boldsymbol{\tilde{b}}$ is a vector of length $n_b$. 

In the LSPG-ROM, the coefficient vector $\cvec(\paramVec)$ is still calculated using Eq.~\eqref{eq:rom_reduction}. 
However, instead of forcing the $\residual(\paramVec)$ in Eq.~\eqref{eq:residual} to be orthogonal to the reduced space, the coefficient vector  $\residual(\paramVec)$ is chosen such that the norm of the residual is minimized~\cite{Quarteroni:218966}, leading to solving the least-squares problem:
\begin{equation}\label{eq:lspg_ls_prob}
    \Amat(\paramVec) \Xmat \cvec(\paramVec) = \bvec(\paramVec) \,.
\end{equation}
The affine decomposition of the residual~\eqref{eq:residual} is given by,
\begin{equation} \label{eq:residual_affine}
\residual(\paramVec)
= \sum_{a=0} \left[  \bvec^{(a)} - \Amat^{(a)} \Xmat \cvec(\paramVec) \right] \paramVec_a \,,
\end{equation}  
where we used the short-hand notation for the column vector $\bvec^{(a)}$ which has components $(\bvec^{(a)})_{i} = \vb{V}_{a, i} = \mathcal{V}_a^{(N+1,i)}$ and the matrix $\vb{A}^{(a)}$ with components $(\vb{A}^{(a)})_{ij}= \{\vb{\mathcal{A}}_{a, ij}\}$.
Equation~\eqref{eq:residual_affine} shows that the residual $\residual$ for \emph{any} value of  $\paramVec$ lies in the space formed by the vectors $\vb{b}^{(a)}$ (length-$N+1$) and the matrix products $\vb{B}^{(a)} = \Amat^{(a)} \Xmat$ ($(N+1) \times n_b$ matrices).
By collecting these terms, we can efficiently build the emulator for \emph{any} parameter value $\paramVec$.

These two sets of vectors are then stacked horizontally to get the $(N+1) \times (n_\theta+1)(n_b+1)$ matrix,
\begin{equation} \label{eq:def_Ymat}
    \Ymat = \begin{bmatrix} \vb{B}^{(0)} & \vb{B}^{(1)} & \cdots & \vb{B}^{(n_\theta)} & \vb{b}^{(0)} & \vb{b}^{(1)} & \cdots & \vb{b}^{(n_\theta)}\end{bmatrix} \,.
\end{equation}
Next, $\Ymat$ is orthonormalized and compressed using a truncated singular value decomposition (SVD), which provides the projection basis. 
This basis has dimensions $N\times n_{\Ymat}$, with $n_{\Ymat} \leqslant (n_\theta+1)(n_b+1)$.
Note that the matrix $\Ymat \Ymat^\dagger$ is constructed to be an orthogonal projector onto the ROM's approximate subspace of residuals.
Finally, this procedure recasts Eq.~\eqref{eq:lspg_ls_prob} to obtain the following emulator equation for the LSPG-ROM:
\begin{multline} \label{eq:lspg_lin_system}
    \tAmat(\paramVec) \cvec(\paramVec) = \tbvec(\paramVec) \\
    \text{with} \quad 
    \tAmat(\paramVec) = \Ymat^\dagger \Amat(\paramVec) \Xmat \,, \quad 
    \tbvec(\paramVec) = \Ymat^\dagger \bvec(\paramVec) \,.
\end{multline}
For more details, we refer to Section~III.B in Ref.~\cite{Maldonado:2025ftg}.

\subsection{Greedy snapshot selection}
\label{sec:greedy_snapshot_selection}

How are the locations of the snapshot calculations determined? 
\rev{Here, we follow the greedy snapshot selection process 
and benchmark it using the proper orthogonal decomposition (POD).} 
We refer the reader to Section~IV in Ref.~\cite{Maldonado:2025ftg} for a comprehensive discussion of these two complementary approaches to snapshot selection.

The greedy algorithm improves upon an initial basis by placing additional snapshots at locations where the emulator error is estimated to be largest.
The initial basis could be chosen randomly, for example, using LHS or in a physics-informed manner.
This active learning method requires robust estimation of emulator errors. 
We use the norm of the residual given in Eq.~\eqref{eq:residual} as a proxy for the true error between the high-fidelity solution vector $\vb{t}(\paramVec)$ and its ROM approximation in Eq.~\eqref{eq:rom_reduction}:
\begin{equation} \label{eq:rom_error}
    \exactErrorVec(\paramVec) = \vb{t}(\paramVec) - \tilde{\vb{t}}(\paramVec)  \,.
\end{equation}
Motivated by the findings in Ref.~\cite{Maldonado:2025ftg}, we assume that $\norm{\rvec(\paramVec)}$ and $\norm{\vb{e}(\paramVec)}$ are approximately proportional to each other, so the factor of proportionality can be determined at the expense of a single additional high-fidelity calculations at the end of the greedy iteration, as discussed in Section~IV.A.3 of Ref.~\cite{Maldonado:2025ftg}.
Under this assumption, the error estimator can hence be calibrated without performing expensive FOM calculations during the emulator's online stage.
A rigorous upper bound of the emulator error, which would constitute a conservative error estimate, was derived in Equation~(48) in Ref.~\cite{Maldonado:2025ftg}. 
However, evaluating it requires a computationally expensive estimate of the minimum singular value of $\vb{A}(\paramVec)$, e.g., using the \rev{successive constraint method}~\cite{HUYNH2007473}.

Furthermore, the residual in the full space~\eqref{eq:residual} can be efficiently computed in terms of the projected residual
\begin{equation}  \label{eq:rom_residual_Y}
    \romResidualY(\paramVec) \equiv \Ymat^\dagger \romResidual(\paramVec) = \Ymat^\dagger \bvec(\paramVec) - \Ymat^\dagger \Amat(\paramVec) \tilde{\vb{t}} (\paramVec) \,,
\end{equation}  
which has $(n_\theta+1)(n_b+1)$ components and, thus, is defined in a semi-reduced space instead of the reduced and full space.
Because $\vb{P} = \Ymat \Ymat^\dagger$ is an orthogonal projector onto the space of the residuals~\eqref{eq:residual}, with $\vb{P} \romResidual(\paramVec) = \romResidual(\paramVec)$ for all $\romResidual(\paramVec)$, the norm of the residual vector in the full space~\eqref{eq:residual} can be obtained exactly and efficiently in the semi-reduced space via
\begin{equation}
     \norm{\romResidual(\paramVec)} = \norm{\romResidualY(\paramVec)} \,.
\end{equation}
In the online stage, we can furthermore reconstruct $\romResidualY(\paramVec)$ from tensors precomputed in the offline stage:
\begin{multline}
\label{eq:res_err_calc}
\boldsymbol{(r_Y)}_w(\boldsymbol{\theta)} = \sum_{a=0}^{n_\theta}\left[\sum_{i=1}^{N+1}\boldsymbol{Y^{\dagger}}_{wi} V_{a,i}\right]_{wa}\boldsymbol{\theta}_a \\
-\sum_{a=0}^{n_\theta}\sum_{u=0}^{n_b}\left[\sum_{i,j = 1}^{N+1} \boldsymbol{Y^{\dagger}}_{wi} \boldsymbol{A}_{a,ij} \boldsymbol{X}_{ju}\right]_{a,wu} \boldsymbol{\theta}_a \boldsymbol{c}_u(\boldsymbol{\theta}) 
\end{multline}
where terms in the bracket are pre-stored. 
And, once $\boldsymbol{r_Y}(\boldsymbol{\theta})$ is reconstructed, its norm is calculated at that $\boldsymbol{\theta}$. 

In the next section, we will discuss how to propagate these emulator errors in the discretized $t$-matrix to scattering cross sections for uncertainty quantification.

\section{Results and Discussion}
\label{sec:results}

In this section, we present our main results for scattering phase shifts and cross sections. 
We begin by analyzing the convergence of the greedy algorithm for the simple Minnesota potential, before discussing the more interesting case of chiral interactions. 

\subsection{Test case: Minnesota potential}

As a test case, we build an emulator for the Minnesota potential~\cite{THOMPSON197753} in the \oneSzero channel, which, in coordinate space, is given by,
\begin{equation} \label{eq:minn_pot_cs}
    V(r, \boldsymbol{\theta}) = V_R \; e^{-\kappa_R r^2} + V_s \; e^{-\kappa_s r^2},
\end{equation}
where $r$ is the relative coordinate.
It was fitted in Ref.~\cite{THOMPSON197753} to match  experimentally extracted effective range parameters, leading to the following best-fit model parameters in Eq.~\eqref{eq:minn_pot_cs}:
$\theta_1\equiv V_R = 200 \MeV$, $\theta_2 \equiv V_s = -91.85 \MeV$, $\kappa_R = 1.487~\text{fm}^{-2}$, and $\kappa_s = 0.465~\text{fm}^{-2}$.
In the following, we will vary the two affine parameters $(V_R, V_s)$ while keeping the non-affine parameters $(\kappa_R,\kappa_s)$ fixed at their best-fit values, and $\theta_0 \equiv 0$.
Because our FOM and ROM solvers are based on the LS equation, we transform Eq.~\eqref{eq:minn_pot_cs} to momentum space by carrying out the Fourier transform analytically, resulting in:\footnote{%
We follow the convention $\Braket{\vb{r}|\vb{p}} = e^{i \vb{r} \cdot \vb{p}} / (2\pi)^{3/2}$.%
}
\begin{multline} \label{eq:minn_pot_ms}
        V(q^2, \paramVec) = 
    \frac{1}{(2\pi)^3} \\ 
     \quad \times \Bigg [ \left(\frac{\pi}{\kappa_R} \right)^{\frac{3}{2}}V_R \; e^{-\frac{q^2}{4\kappa_R}} + \left(\frac{\pi}{\kappa_s}\right)^{\frac{3}{2}}V_s \; e^{-\frac{q^2}{4\kappa_s}} \Bigg],
\end{multline}
with the momentum transfer squared $q^2 = |\vb{p}'-\vb{p}|^2$ and the initial and final relative nucleon momenta $\vb{p}$ and $\vb{p}'$, respectively.
Note that the Fourier transform is linear, so Eqs.~\eqref{eq:minn_pot_cs} and~\eqref{eq:minn_pot_ms} have similar affine decompositions.

\begin{figure*}[tb]
    \centering
    \includegraphics[width=\textwidth]{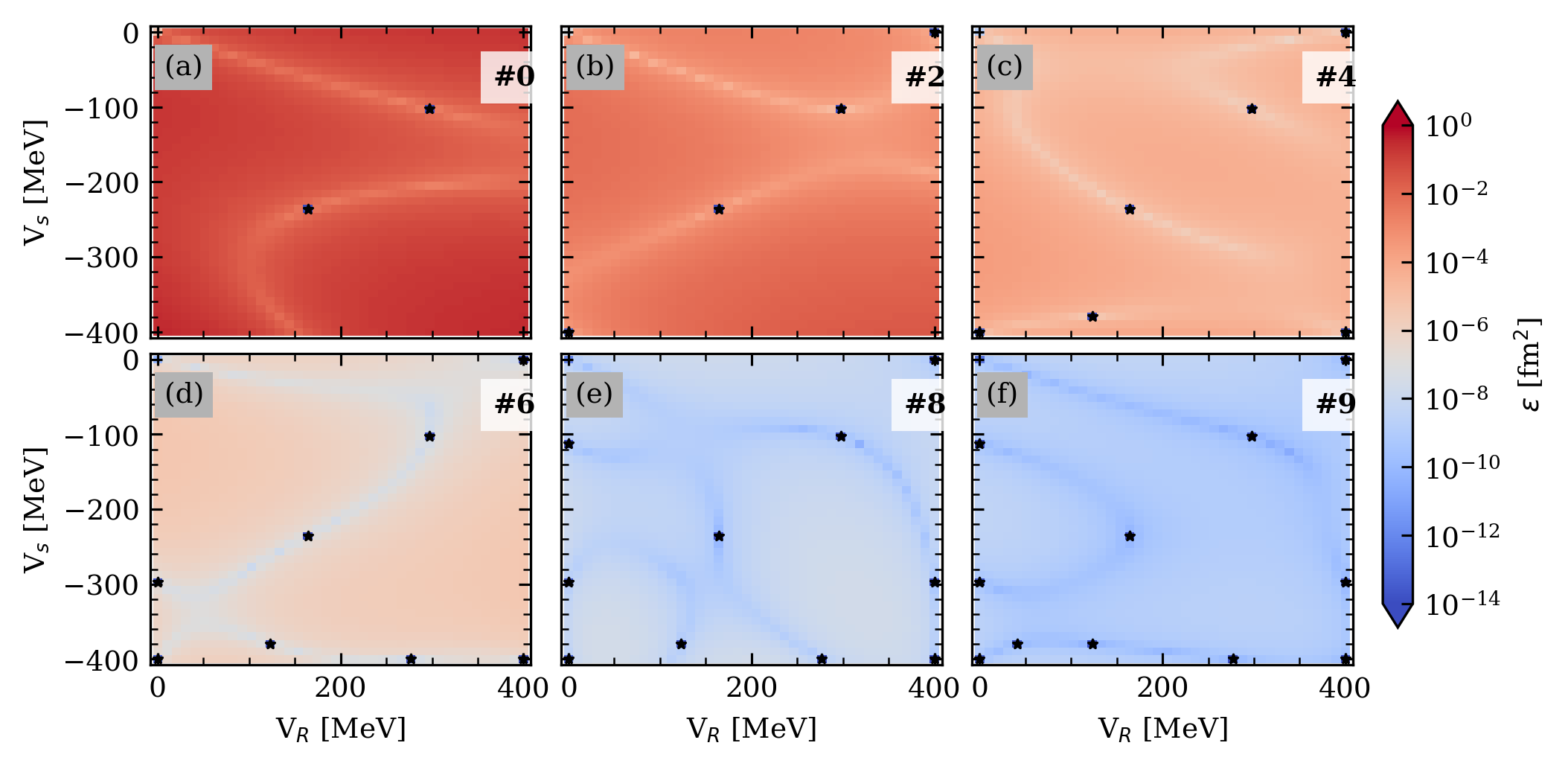}
    \caption{%
    Convergence of the greedy algorithm for the Minnesota potential~\eqref{eq:minn_pot_ms} at $E_\text{lab} = 20 \MeV$ in the \oneSzero channel.
    Each panel shows a color map of the true error~\eqref{eq:rom_error} in the two-dimensional parameter space of the potential (see the legends).
    The pound symbol ``\#'' specifies the number of greedy iterations passed: 
    Panel~(a) depicts the initial configuration, for which we randomly select snapshots in the parameter space using LHS. 
    The remaining panels~(b) through~(f) show the configuration in the emulator after 2, 4, 6, 8, and 9 iterations, respectively. 
    Note that not all greedy iterations are depicted.
    As shown, the greedy algorithm places additional snapshots at locations where the (estimated) emulator error is largest, thereby iteratively reducing the emulator error locally and globally.
    Note that the emulator error vanishes at the origin (which is not a grid point in the panels) because the interaction is zero as there is no parameter-independent term, rendering the ROM equations homogeneous. %
    }
    \label{fig:2D_minnesota_greedy}
\end{figure*}

Figure~$\ref{fig:2D_minnesota_greedy}$ shows the greedy algorithm in action for the Minnesota potential~\eqref{eq:minn_pot_ms} at a representative laboratory energy, $E_\text{lab}= 20 \MeV$.%
\footnote{See also Figure~6 in Ref.~\cite{Maldonado:2025ftg} for a similar, albeit one-dimensional, illustration of the greedy algorithm in action.}
Each panel in Fig.~\ref{fig:2D_minnesota_greedy} depicts a color map of the true error~\eqref{eq:rom_error} in the two-dimensional parameter space of the potential (see the legends).
Specifically, Fig.~\ref{fig:2D_minnesota_greedy}a) shows the initial configuration, for which we randomly select two snapshots in the parameter space using LHS. 
The black squares near the center of the shown parameter space highlight the snapshot location.
At these points, the emulator error vanishes, as expected.

From this initial configuration, we let the greedy algorithm improve the snapshot basis by adding one snapshot per iteration:
Panels~b) through~f) show the color map of the actual error after 2, 4, 6, 8, and 9 of those iterations, respectively, as indicated by the number following the pound symbol (``\#'') in the annotated text. 
Note that, for brevity, not all iterations are depicted in Fig.~\ref{fig:2D_minnesota_greedy}.
\rev{At the end of each iteration, we (re)orthonormalize the emulator basis by applying the QR decomposition to the snapshot matrix $\Xmat$.} 
An orthonormal emulator basis significantly improves the numerical stability of the ROM.
The greedy algorithm terminates after nine iterations (see panel~f) because all errors shown are below the requested threshold, here set to $\varepsilon < 10^{-8}$.  
After these iterations, the emulator basis consists of $2+9=11$ snapshots (corresponding to the dark points in panel~f).

As illustrated by Fig.~\ref{fig:2D_minnesota_greedy}, the greedy algorithm enhances a given snapshot basis by placing additional snapshots in locations where the estimated emulator error is the largest. 
The estimated location of maximum error in this application closely matches the actual location. 
The greedy algorithm thus effectively reduces the emulator error locally and globally until the requested tolerance $\tol$ is met.
Furthermore, we observe in Fig.~\ref{fig:2D_minnesota_greedy} that the greedy algorithm adds snapshot calculations at the boundaries of the parameter space to the emulator basis.
This behavior was already observed in Ref.~\cite{Maldonado:2025ftg}.
In these regions, the emulator must extrapolate so that larger errors can be expected.
While in two-dimensional parameter spaces one can manually place snapshot locations at the boundaries to improve the emulator's accuracy, in higher-dimensional spaces this may no longer be possible due to the curse of dimensionality. 
On the other hand, the greedy algorithm always places snapshots so that the emulator error is optimally reduced locally, including choosing snapshots at the boundaries that are most important for error reduction. 

\subsection{Chiral potential at \texorpdfstring{\nTwoLO}{N2LO}}
\label{sec:chiral_pot}

Next, we apply the greedy emulators to a realistic chiral potential. 
Following Refs.~\cite{Drischler:2021qoy,Maldonado:2025ftg}, we choose here one of the local chiral potentials constructed by Gezerlis~et al.\ in Ref.~\cite{Gezerlis:2014zia}: 
the one in the $np$-channel at \nTwoLO and with coordinate space cutoff $R = 1.0 \fm$ and spectral function cutoff $\tilde{\Lambda} = 1000 \MeV$.
We will refer to this potential as the ``GT+ potential.''
These local chiral potentials are commonly used in quantum Monte Carlo calculations of finite nuclei and infinite matter~\cite{Piarulli:2019cqu}.
However, we note that the GT+ potential is merely a convenient choice and that our FOM and ROMs are broadly applicable to any local or non-local potential in momentum space \rev{for which an affine decomposition~\eqref{eq:affine_potential} is possible.}
The affine parameters of the GT+ potential are the NN contact LECs $\paramVec =  \{1, C_S, C_T, C_1, C_2, ..., C_7\}$, where we added again an auxiliary dimension to incorporate constant contributions; here, all pionic terms are kept at their best-fit values. 
The authors of Ref.~\cite{Gezerlis:2014zia} provide a C$^{++}$ code that evaluates their potentials in both coordinate and momentum space. 
We modified this code to output the affine components $\mathcal{V}_a^{(\ell,\ell')}(k,k')$ in Eq.~\eqref{eq:affine_potential} in momentum space.
In the following, we consider variations in all nine LECs to benchmark our emulators, although not all of them contribute to all partial-wave channels, as shown in Appendix~I of Ref.~\cite{Gezerlis:2014zia}.

\subsubsection{Emulator convergence}
\label{sec:emu_conv}

We study the convergence of the greedy algorithm and benchmark it against the POD approach to snapshot selection. 
For emulator training, we use a set of 150 random sampling points obtained via LHS in the region of $\pm 40 \%$ of each LEC's best-fit value~\cite{Gezerlis:2014zia}.
Although both have access to the same candidate snapshots, the greedy and POD approaches handle them differently during training.
As illustrated in Fig.~1 in Ref.~\cite{Maldonado:2025ftg}, the POD approach computes high-fidelity solutions at all candidate snapshots and then compresses the emulator basis using a truncated SVD. 
In contrast, the greedy emulator performs high-fidelity calculations only on a small subset of candidate snapshots, namely those for which the emulator error was estimated to be largest during the greedy iteration.
The emulator error can be estimated efficiently without requiring high-fidelity calculations, as discussed in Sec.~\ref{sec:greedy_snapshot_selection}.
For more details, see Section IV.B in Ref.~\cite{Maldonado:2025ftg}.
After training, we validate the emulators using 500 random points obtained via LHS similar to the candidate snapshots.

\begin{figure*}[tb]
    \centering
    \includegraphics[width=\textwidth]{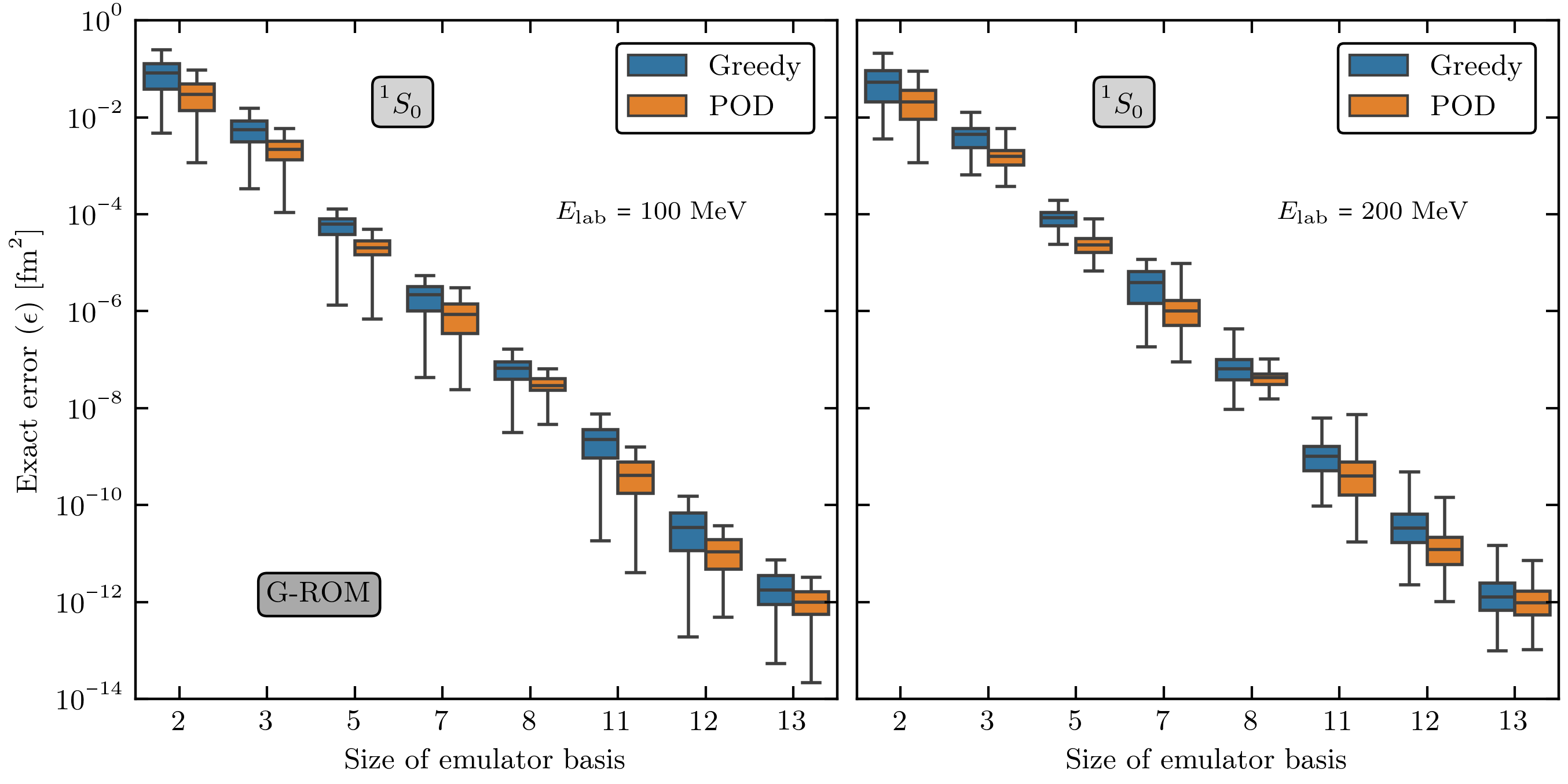}
    \caption{%
    Emulator convergence and comparison at $E_\text{lab} = 100 \MeV$ (left panel) and $200 \MeV$ (right panel) for the GT+ potential~\cite{Gezerlis:2014zia} in the \oneSzero channel. 
    The greedy algorithm applied to the G-ROM is compared to the POD approach. 
    Both emulators have access to the 150 candidate snapshots in a nine-dimensional parameter space, randomly selected via LHS in the region of $\pm 40\%$ of the best-fit LEC values (obtained in Ref.~\cite{Gezerlis:2014zia}) 
    However, the POD emulator uses all of them, while the greedy emulator selects only a small subset.
    The $x$-axis shows the size of the emulator basis, corresponding to the snapshot calculations performed (greedy algorithm) or the dominant POD modes used (POD approach).
    To test their accuracy, we use a validation set of 500, similarly obtained via LHS as the candidate snapshots during training.
    The boxes indicate the range between the first and third quartiles, and the line within the box represents the median. 
    The whiskers show the range from the 5th to the 95th percentiles.%
    }
    \label{fig:singlet_error}
\end{figure*}

\begin{figure*}[tb]
    \centering
    \includegraphics[width=\textwidth]{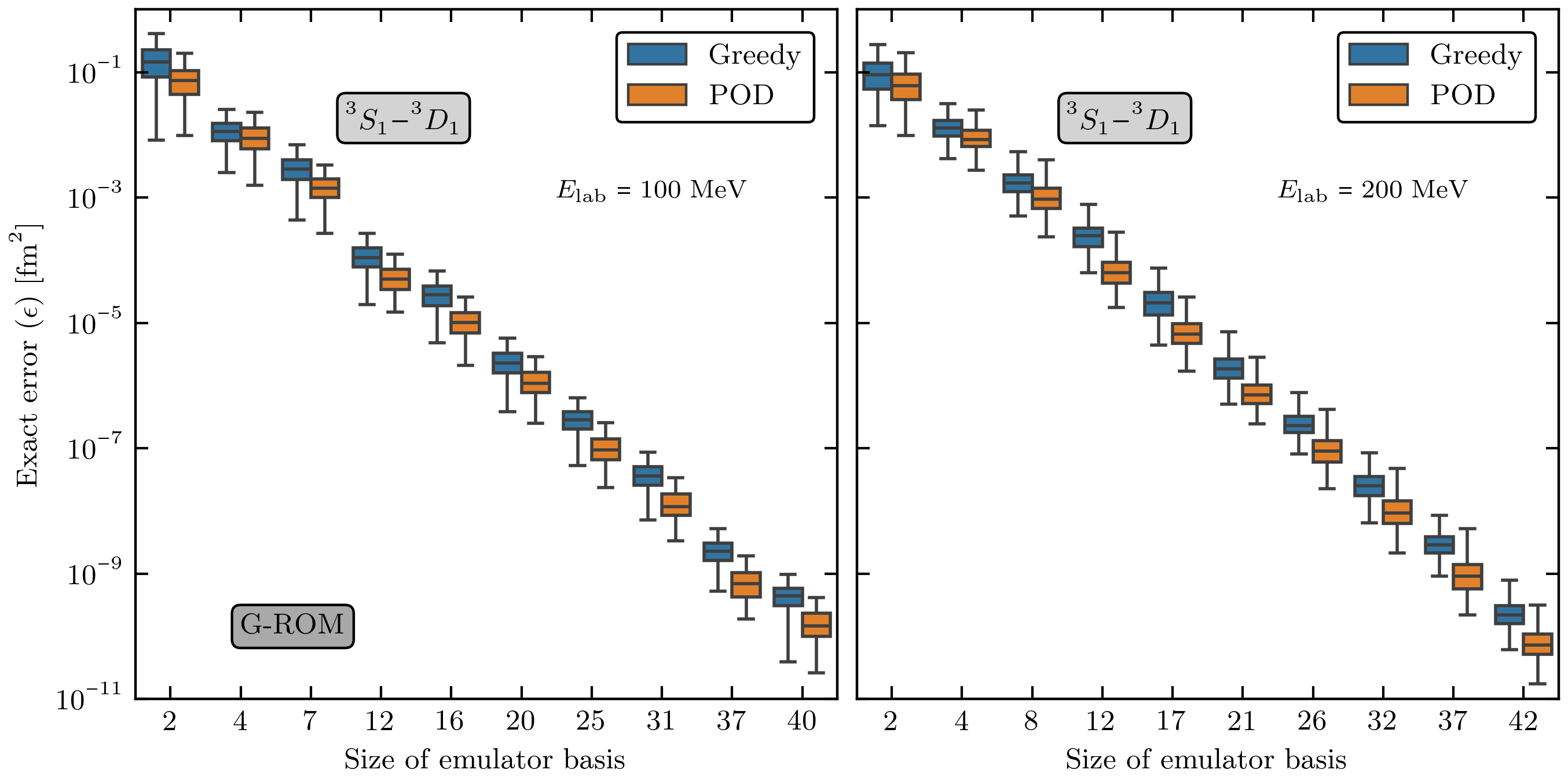}
    \caption{Same as Fig.~\ref{fig:singlet_error} but for the coupled channel \threeSDone.}
    \label{fig:coupled_error}
\end{figure*}

Figures~\ref{fig:singlet_error} and~\ref{fig:coupled_error} show, using so-called box-and-whisker plots, the distributions of the exact relative errors~\eqref{eq:rom_error} in the uncoupled \oneSzero and coupled \threeSDone channels, respectively, based on the validation points.\footnote{%
See also Figures~7 through~10 in Ref.~\cite{Maldonado:2025ftg} for similar convergence plots in the \oneSzero channel.%
}
The left panels correspond to $E_\text{lab} = 100 \MeV$, the right panels to $E_\text{lab} = 200 \MeV$.
We observe similar convergence patterns for the two ROMs and thus focus here on the G-ROM. 
The filled boxes indicate the range from the first to the third quartiles, and the line within the box represents the median. 
In addition, the whiskers show the range from the 5th to the 95th percentiles.
For the greedy approach, the $x$-axis specifies the number of snapshot calculations performed in total, whereas for the POD approach, it specifies the number of dominant POD modes.
In both cases, it specifies the size of the emulator basis.

For both partial-wave channels and energies, we observe that the medians of the relative error~\eqref{eq:sym_relative_error} decrease exponentially as the emulator basis size increases.
The greedy and POD approaches consistently produce accurate and precise predictions for
both partial-wave channels.
For the \oneSzero channel in Fig.~\ref{fig:singlet_error}, we observe that $n_b \gtrsim 7$ is required to obtain an exact error of $\varepsilon \lesssim 10^{-6}$, independent of the two energies shown.
For the \threeSDone channel in Fig.~\ref{fig:coupled_error}, achieving the same exact error requires about $n_b \gtrsim 20$ for both energies.
These basis sizes are significantly smaller than the number of candidate snapshots.

As in Ref.~\cite{Maldonado:2025ftg}, we find that the greedy algorithm achieves similar accuracies than the POD, although its errors are systematically slightly larger.
This observation is not surprising because the POD approach has access to the most information and uses, by construction, the most important emulator basis vectors in terms of the singular values (associated with the dominant POD modes).
However, the POD approach has two disadvantages over the greedy approach: 
First, it requires far more high-fidelity calculations than the other emulators, which may not be affordable in computationally more demanding applications, such as three-body scattering (see Ref.~\cite{Gnech:2025gsy,Gnech:2025lbg}).
Second, it does not estimate errors.

We also investigate how well our emulators preserve (or violate) the unitarity of the scattering $S$-matrix. 
In all applications with sufficiently small requested emulator tolerances, i.e., $\tol \gtrsim 10^{-2}$, we find that unitarity violations are insignificant.
Specifically, for uncoupled channels, we observe that unitarity is generally preserved at machine precision, whereas for the coupled channels, it is of the order of the emulator error.
implying that it decreases as the emulator basis improves during the greedy algorithm. 
Hence, the greedy algorithm systematically remedies violations of unitarity until a desired threshold is achieved.

\subsubsection{Emulation of phase shifts}

\begin{figure*}[htb]
    \centering
    \includegraphics[width=\textwidth]{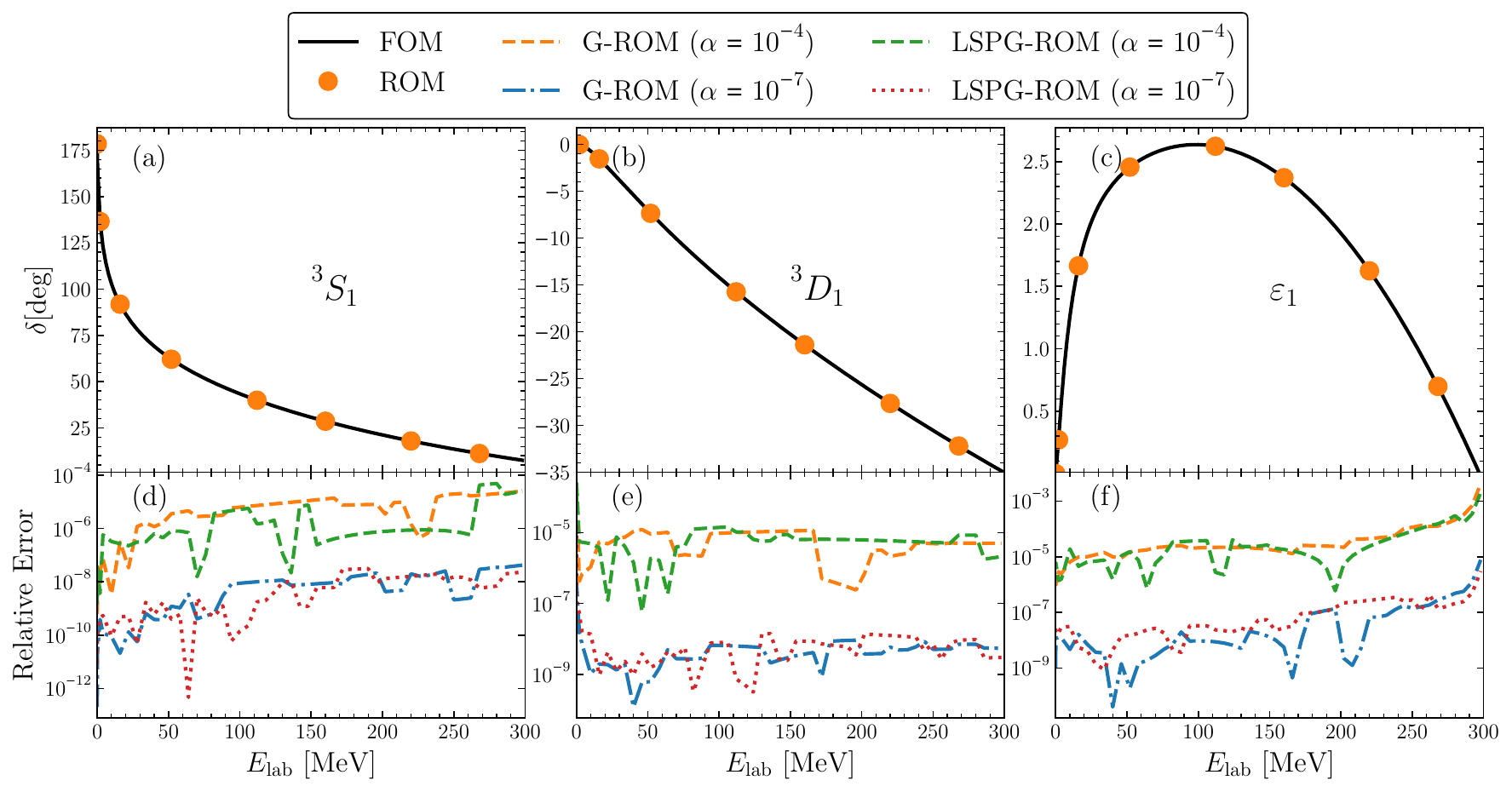}
    \caption{%
    High-fidelity (or FOM) and emulated (or ROM) phase shifts and mixing angles as a function of the laboratory energy for the coupled \threeSDone channel based on the GT+ potential. 
    The panels~(a) to~(c) show the phase shifts in the \threeSone (panel~a) and in the \threeDone channel (panel~b), as well as the mixing angle 
    $\varepsilon_1$ (panel~c). 
    The solid black lines represent the high-fidelity results, while the dots indicate the emulated results at a few selected energies. 
    The panels~(d) to~(f) show the relative error~\eqref{eq:sym_relative_error} of the G-ROM and LSPG-ROM, respectively, with respect to the high-fidelity calculations.
    This comparison was made for two requested error tolerances on the $\tilde{t}$-matrix, $\tol = 10^{-4}$ and $10^{-7}$, as reported in the legend. %
    }
    \label{fig:coupled_phases}
\end{figure*}

Next, we emulate phase shifts based on the GT+ potential. 
To this end, we construct emulators on a grid of laboratory energies for a given partial-wave channel.
Figure~\ref{fig:coupled_phases} shows our results for the emulated phase shifts and mixing angles at the potential's best-fit values of the LECs as a function of the laboratory energy in the coupled \threeSDone channel.\footnote{%
The best-fit values were obtained in Ref.~\cite{Gezerlis:2014zia} using a least-squares fit to scattering phase shifts and the \oneSzero scattering lengths.%
}
We choose the G-ROM and the deuteron channel as a coupled-channel case. 
The results are representative of the LSPG-ROM and other partial-wave channels we have studied, including the coupled and uncoupled P and F partial waves.\footnote{%
Note that the approach in Ref.~\cite{Maldonado:2025ftg} was limited to uncoupled channels and local potentials (in coordinate space), restrictions that our emulators relax.%
}

To obtain the emulated phase shifts in Fig.~\ref{fig:coupled_phases}, we train the emulator using a pool of 500 candidate snapshots randomly selected in the region of $\pm 40 \%$ of the best-fit LECs values obtained in Ref.~\cite{Gezerlis:2014zia}. 
Given a single initial snapshot randomly chosen from this pool, the greedy algorithm estimates the error of each remaining candidate snapshot, performs a single high-fidelity calculation for the one with the largest (estimated) error, and adds the orthonormalized snapshot to the emulator basis. 
This process is repeated until the requested error tolerance $\tol$ specified in the legend is met.
After training is complete, we run the emulators using the best-fit LEC values from Ref.~\cite{Gezerlis:2014zia}.

Specifically, the panels~(a), (b), and~(c) in Fig.~\ref{fig:coupled_phases} show, from left to right, the phase shifts in the \threeSone and \threeDone channels, and the corresponding mixing parameter $\varepsilon_1$. 
The solid lines depict the high-fidelity (or FOM) calculations, and the filled circles represent the emulated (or ROM) results at a few selected energies. 
The panels~(d), (e), and~(f) in Fig.~\ref{fig:coupled_phases} depict, for each ROM, the corresponding symmetric relative error with respect to emulator predictions:
\begin{equation} \label{eq:sym_relative_error}
    e(x,\tilde{x}) = 2 \frac{|x-\tilde{x}|}{|x|+|\tilde{x}|}  \,,
\end{equation}
where the emulated and high-fidelity results are denoted by $\tilde{x}$ and $x$, respectively.

To assess how well our emulator approximates phase shifts, we train two G-ROM emulators with different requested tolerances, $\tol = 10^{-4}$ (orange and green lines) and $10^{-7}$ (blue and red lines). 
Note that these tolerances are requested for the emulated $t$-matrix in discretized form, not on the phase shifts directly.
With $\tol = 10^{-4}$, both G-ROMs give phase shifts and mixing angles with mean relative errors of $\lesssim 10^{-5}$, which decreases to $\lesssim 10^{-7}$ with $\tol = 10^{-7}$.

We conclude that the greedy algorithm enables efficient, accurate training of scattering emulators. 
Note that second-order corrections from applying the \rev{Kohn variational principle}~\cite{Furnstahl:2020abp,Garcia:2023slj} are not needed to achieve this level of accuracy.

In contrast to Ref.~\cite{Maldonado:2025ftg}, we have not encountered spurious singularities known as Kohn (or Schwartz) anomalies that occur when the emulator equations are singular or near-singular~\cite{Drischler:2021qoy}.
This may suggest that scattering calculations in momentum space are less prone to such anomalies, regardless of the ROM used, likely because the $t$-matrix is complex-valued, as discussed in Ref.~\cite{Drischler:2021qoy}. 
Only the LSPG-ROM has been analytically shown in Ref.~\cite{Maldonado:2025ftg} to be less prone to Kohn anomalies. 
  
\subsubsection{Emulation of total cross sections}
\label{sec:emu_scatt_obs}

\begin{figure*}[tb]
    \centering
    \includegraphics[width=\textwidth]{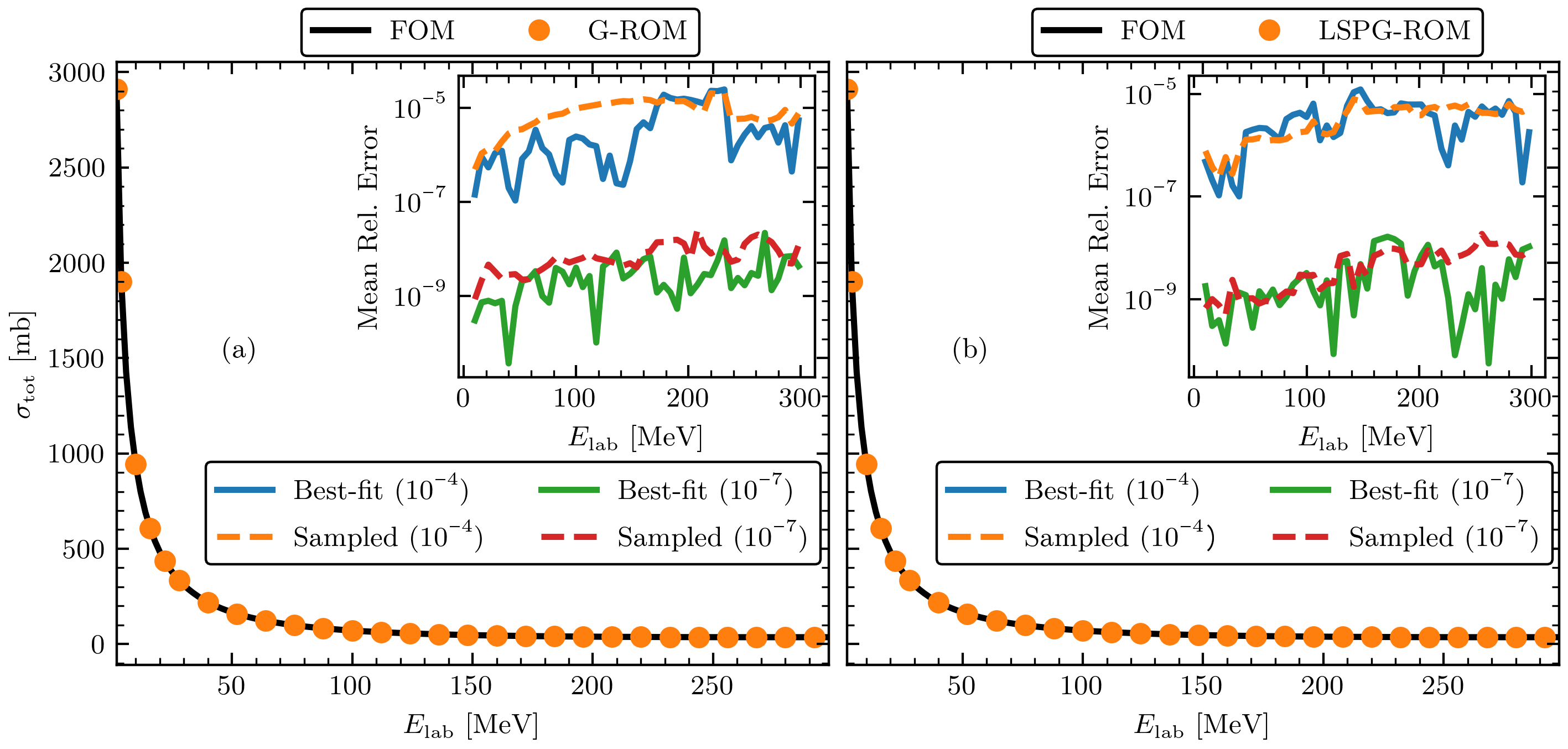}
    \caption{%
    Total $np$ cross sections obtained with the G-ROM in panel~(a) and LSPG-ROM in panel~(b) (orange dots) based on the GT+ potential, which depends on nine LECs.
    The black lines depict the high-fidelity calculations.
    All partial-wave channels with $j \leqslant 6$ are included in the calculations. 
    Both emulators use the same pool of candidate snapshots for training, comprising 500 random points selected via LHS.
    The insets show the mean relative error~\eqref{eq:sym_relative_error} of the emulator and the high-fidelity solution based on 1000 random samples in the potential's parameter space (dashed lines) and the best-fit LEC values obtained in Ref.~\cite{Gezerlis:2014zia} (solid lines), for the requested error tolerances of $\tol = 10^{-4}$ and $10^{-7}$ (see the legends). %
    }
    \label{fig:coupled_cross}
\end{figure*}

Next, we benchmark our emulators by calculating the $np$ total cross section~\eqref{eq:tot_cs} as a function of the laboratory energy, using the GT+ potential. 
Figure~\ref{fig:coupled_cross} compares the emulator predictions (orange dots) and high-fidelity calculations (black lines).
The results for the G-ROM are depicted in panel~(a); for the LSPG-ROM in panel~(b).
All partial-wave channels with $j \leqslant 6$, coupled and uncoupled, are included in these calculations. 
The insets show the mean value of the relative error~\eqref{eq:sym_relative_error} of the emulators based on 1000 random samples in the potential's nine-dimensional parameter space (dashed lines) and the best-fit LEC values obtained in Ref.~\cite{Gezerlis:2014zia} (solid lines). 
For emulator training, we generate a pool of 500 candidate snapshots, randomly selected via LHS within $\pm 40 \%$ of the best-fit LEC values reported in Ref.~\cite{Gezerlis:2014zia}.
Both ROMs have access to the same candidate snapshots during training.

For this emulator comparison in Fig.~\ref{fig:coupled_cross}, we again consider the two requested error tolerances  $\tol = 10^{-4}$ and $10^{-7}$ (see the legends).
Note that this tolerance is requested for the error in the discretized $t$-matrix (i.e., the vector $\vb{t}$), not in the total cross section shown in Fig.~\ref{fig:coupled_cross}. 
We will discuss the propagation of this error to the total cross section in Sec.~\ref{sec:stat_framework}.
However, it may not be surprising that the relative error~\eqref{eq:sym_relative_error} in the total cross section also decreases by approximately three orders of magnitude, from $\approx 10^{-6}$ to $\approx 10^{-9}$, if the requested error tolerances is lowered from $10^{-4}$ to $10^{-7}$ in the energy range depicted in Fig.~\ref{fig:coupled_cross}.

Overall, we observe that both emulators accurately reproduce the cross sections up to $E_{\text{lab}} = 300 \MeV$, validating that they can reliably reproduce total cross sections, not only at the best-fit LEC values but also across the broader parameter space sampled. 

\subsection{Proof of principle: Bayesian parameter estimation with emulator errors}

Next, we explore the efficacy of our emulators incorporating error estimation for Bayesian calibration of chiral NN interactions.
We perform these calculations here only as a proof-of-principle step toward future, more rigorous, applications in sophisticated Bayesian frameworks, such as those in Refs.~\cite{Melendez:2017phj,Melendez:2019izc,Wesolowski:2021cni,Svensson:2022kkj,Svensson:2023twt}.
Our goal is to demonstrate that emulated errors for scattering observables can be estimated and accounted for in Bayesian calibration, and to assess the emulators' computational efficiency.

To this end, we revisit the GT+ potential obtained in Ref.~\cite{Gezerlis:2014zia} and constrain its NN contact interactions using total cross-section data from the partial-wave analysis in Ref.~\cite{Stoks:1993tb}.
However, let us emphasize again that our emulators are also applicable to more modern chiral NN interactions, including the LENPIC SMS potentials~\cite{Reinert:2017usi,Epelbaum:2022cyo}.
Note that total cross sections are direct observables in contrast to the phase shifts.

\subsubsection{Statistical framework}
\label{sec:stat_framework}

Our Bayesian parameter estimation is based on the statistical discrepancy model:
\begin{equation} \label{eq:discrepancy_model}
    y_\text{exp}(x) = y_\text{th}(x; \paramVec) + \delta y_\text{th}(x; \paramVec) + \delta y_\text{exp}(x) \,, 
\end{equation}
where $y(x)$ stands for the observable of interest, here the total cross section $y = \sigma_\text{tot}$ as a function of the energy $x = E$, and $\delta y(x)$ stands for its uncertainty.
The subscripts refer to the experimentally informed values (``exp'') and the theoretical predictions (``theo'').
For the theory uncertainties, 
\begin{equation} \label{eq:total_th_err}
    \delta y_\text{th}(x; \paramVec) = \delta y_{\text{emu}}(x; \paramVec) + \delta y_{\text{EFT}}(x) \,,
\end{equation}
we include the emulator error and assume the point-wise EFT truncation error model developed by the BUQEYE collaboration~\cite{Melendez:2020xcs}. 
We further assume that these uncertainties are independent and that the theory uncertainties dominate the uncertainty budget, i.e., $\delta y_\text{exp}(x) \ll \delta y_\text{th}(x; \paramVec)$.
Future applications will more rigorously estimate the theory uncertainties, including correlated EFT truncation errors following the prescription in Ref.~\cite{Melendez:2019izc}.

To estimate the emulator error $y_\text{emu}(x)$ from the estimated error in the discretized $t$-matrix (i.e., the vector $\tilde{\vb{t}}$), we apply the triangle inequality repeatedly to Eq.~\eqref{eq:tot_cs}, leading to the upper bound for the error in the emulated cross section as follows:
\begin{subequations} \label{eq:yemu_cs}
    \begin{align}
    &\varepsilon_\text{emu}(x; \paramVec) \equiv \norm{\sigma_{\text{tot}} - \widetilde{\sigma}_{\text{tot}}} \\
    & \quad \leqslant \frac{\pi}{k_0^2} \sum_{j} (2j+1) \sum_{k} \norm{\tau^j_{k} (E_{k_0}) -\widetilde{\tau}^j_{k} (E_{k_0}) } \\
    & \quad \lesssim \frac{\pi^2}{k_0} \mu \sum_{j} (2j+1) \sum_{k} \norm{\exactErrorVec_k}  \,.
\end{align}
\end{subequations}
The summation index $k$ runs over the coupled and uncoupled channels associated with a given total angular momentum $j$.
To obtain Eq.~\eqref{eq:yemu_cs}, we assumed the conservative case in which the emulated on-shell component in $\vb{t}$ has maximum error, while the off-shell components (which do not contribute to $\sigma_\text{tot}$) have no error.
As described in Sec.~\ref{sec:greedy_snapshot_selection}, we also assume that $\norm{\exactErrorVec_k} \propto \norm{\rvec_k}$ and determine the factor of (approximate) proportionality for calibration as described in Ref.~\cite{Maldonado:2025ftg}. 

Since Eq.~\eqref{eq:yemu_cs} provides an (approximate) upper bound on the true emulator error without preferring values inside the error bound, we assume that $\delta y_\text{emu}(x; \paramVec)  > 0$ is point-wise uniformly distributed:
\begin{equation}
    \delta y_{\text{emu}}(x; \paramVec) \sim \mathcal{U} \left(-\varepsilon_\text{emu}(x; \paramVec), \varepsilon_\text{emu}(x; \paramVec) \right) \,.
\end{equation}

We follow the point-wise truncation error model laid out in Appendix~A in Ref.~\cite{Melendez:2019izc} to estimate the to-all-orders EFT truncation error, 
\begin{subequations} \label{eq:eft_trunc_error}
    \begin{align}
\delta y_k(x) &= y_\text{ref} (x) \sum \limits_{n=k+1}^\infty c_n(x)Q^n(x) \,,\\
    c_n \given \bar{c}^2 &\sim \mathcal{N}(0, \bar{c}^2) \,,\\
\bar{c}^2 &\sim \chi^{-2}(\nu_0, \tau_0^2) \,,
\end{align}
\end{subequations}
associated with the EFT prediction at order $k$,
\begin{equation}
    y_\text{th}(x) \equiv y_k(x) = y_\text{ref} (x) \sum \limits_{n=0}^k c_n(x) Q^n(x) \,.
\end{equation}
We assume that the reference scale $y_\text{ref}(x)$ is given by the EFT prediction at the best-fit values  $\paramVec^*$ obtained in Ref.~\cite{Gezerlis:2014zia}, $y_\text{ref}(x) = y_\text{th}(x, \paramVec^*)$.
At low energies, we can further assume that the EFT expansion parameter amounts to $Q = m_{\pi} / \Lambda_b$, where $m_{\pi} \approx 140 \MeV$ is the average pion mass and $\Lambda_b \approx 600 \MeV$ is the EFT breakdown scale~\cite{Melendez:2019izc}.
Note that the $c_n(x)$  are not the LECs of the underlying interactions.
In Eqs.~\eqref{eq:eft_trunc_error}, $\chi^{-2}(\nu_0, \tau_0^2)$ denotes the inverse-chi-squared distribution, with $\nu_0$ degrees of freedom and scale parameter $\tau_0^2$. 
As derived in Refs.~\cite{Melendez:2017phj,Melendez:2019izc}, the EFT truncation error~\eqref{eq:eft_trunc_error} conditional on $\bar{c}^2$ and $Q$ can then given by the normal distribution:
\begin{equation} \label{eq:delta_yk_eft}
    \delta y_k \given \bar{c}^2, Q \sim \mathcal{N} \left[ 0, y_\text{ref}^2(x) \frac{Q^{2(k+1)}}{1-Q^2} \bar{c}^2\right] \,.
\end{equation}
At \nTwoLO ($k=3$), one therefore finds for the variance of the normal distribution~\eqref{eq:delta_yk_eft}
\begin{equation} \label{eq:eft_err_sigma}
    \sigma_k^2(x) \approx \left(0.003\bar{c}\right)^2 \times y_\text{th}^2(x, \paramVec^*) \,,
\end{equation}
indicating that the EFT truncation error is a fraction of $y_\text{th}^2(x, \paramVec^*) $.
Here, we choose a realistic $\bar{c} = 1$~\cite{Millican:2025sdp,Wesolowski:2021cni,Melendez:2019izc}.

Applying the discrepancy model~\eqref{eq:discrepancy_model} together with Bayes' theorem allows us to obtain the posterior distribution for the model parameters,
\begin{equation} \label{eq:posterior}
    \pr( \vb*{\theta} \given \mathcal{D}) = 
    \frac{\pr(\mathcal{D} \given \vb*{\theta}) \pr( \vb*{\theta})}{\pr( \mathcal{D})}\,,
\end{equation}
for the nine NN contact LECs contributing at \nTwoLO.
Here, the measured cross section data is denoted by $ \mathcal{D} = \{(E_n,\sigma_n)\}_n$,  
the likelihood of the data given the LECs $ \vb*{\theta}$ by $\pr(\mathcal{D} \given \vb*{\theta})$, 
the prior distribution encoding our prior knowledge of the LEC distribution by $\pr(\vb*{\theta})$, 
and the Bayesian evidence by $\pr (\mathcal{D})$.
In what follows, we will describe these probability distributions in more detail.

Assuming the EFT and emulator errors are independent of one another, the likelihood for a \emph{single} calculation at $x$ is given by the convolution integral:\footnote{%
For brevity, we drop the subscript ``exp'' that indicates the experimentally informed value in the discrepancy model~\eqref{eq:discrepancy_model} from here on.%
}
    \begin{align}
\pr(y \given \paramVec) &= 
\int \dd{(\delta y_{\text{emu}})}  \pr(y-\delta y_{\text{emu}}) \; \pr(\delta y_{\text{emu}})  \,,\\
\begin{split} 
    &=
    \frac{1}{2 \varepsilon_\text{emu}(x; \paramVec)}
    \Biggl[
        \Phi \left( \frac{d(x,\paramVec) + \varepsilon_\text{emu}(x; \paramVec)}{\sigma_k(x)} \right)\\
        &\quad -
        \Phi \left(\frac{d(x,\paramVec) - \varepsilon_\text{emu}(x; \paramVec)}{\sigma_k(x)}\right)
    \Biggr] \,,
\end{split} \label{eq:single_likelihood}
\end{align}
with the cumulative distribution function,
\begin{align}
    \Phi(z) &= \frac{1}{\sqrt{2\pi}} \int_{-\infty}^{z} \exp\left(-\frac{t^2}{2}\right)\ \dd{t}\,,
\intertext{the residual}
    d(x,\paramVec) &= y(x) - y_{\text{th}}(x;\paramVec) \,,
\intertext{and the normal distribution}
    \pr(y-\delta y_{\text{emu}}) &= \mathcal{N} \left(
       y - \delta y_{\text{emu}} \given y_{\text{th}}(x;\paramVec) , \sigma_k^2(x)  
    \right) \,.
\end{align}
As expected, in the limit of vanishing emulator errors, the likelihood~\eqref{eq:single_likelihood} reduces to the prediction with the point-wise EFT truncation error only:
\begin{equation}
     \lim_{\delta y_\text{emu}(x,\paramVec)\to 0} \pr(y \given  \paramVec)
    =
    \mathcal{N}\left(
        y \given y_{\text{th}}(x;\paramVec), \sigma_k^2
    \right)\,.
\end{equation}
Assuming point-wise errors without correlations across energy, one arrives at the joint log-likelihood function:
\begin{equation} \label{eq:log_likelihood}
\begin{split}
    \ln \pr(\mathbf{y}\given \paramVec)
    &=
    \sum_{n}
    \Biggl[
        -\ln\left(2 \varepsilon_\text{emu}(x_n; \paramVec)\right)\\
        &\quad +
        \ln\biggl(
            \Phi\left(\frac{d(x_n,\paramVec) + \varepsilon_\text{emu}(x_n; \paramVec)}{\sigma_k(x_n)}\right)
            \\ &\quad -
            \Phi\left(\frac{d(x_n,\paramVec) - \varepsilon_\text{emu}(x_n; \paramVec)}{\sigma_k(x_n)}\right)
        \biggr)
    \Biggr]\,.
\end{split}
\end{equation}

The LEC prior distribution is taken to be an uncorrelated multivariate normal distribution with mean vector set to the best-fit values $\paramVec^*$ obtained in Ref.~\cite{Gezerlis:2014zia}:
\begin{equation} \label{eq:prior}
    \pr(\paramVec) = \mathcal{N} \left(
            \paramVec \given \paramVec^*, \operatorname{diag}(\kappa \paramVec^*)^2
    \right)\,,
\end{equation}
with an arbitrarily chosen scaling factor $\kappa = 0.1$ characterizing its width.
The choices we have made for this proof-of-principle calculation can be straightforwardly adjusted using our source codes available on GitHub~\cite{BUQEYEsoftware}.

Finally, the Bayesian evidence in the posterior~\eqref{eq:posterior} serves here only as a normalization constant and can thus be ignored.
However, computing the evidence would be interesting for comparing different EFT models using Bayes factors in future work.

We use the Python library \texttt{emcee} to perform Markov chain Monte Carlo (MCMC) sampling of the posterior~\eqref{eq:posterior}.
The initial positions of the Markov chains are set to the maximum a posteriori (MAP) estimate plus small perturbations. 
We obtain the MAP value by minimizing the negative log-posterior using a two-step procedure: 
First, a global basin-hopping search with 50 iterations and L-BFGS-B as the local optimizer, and then a final L-BFGS-B minimization starting from the best basin-hopping result. 
We then initialize 32 walkers scattered around this MAP point and run \texttt{emcee} for 5000 warm-up steps. 
After this warm-up, we reset the sampler and run an additional 20000 steps, which we use as our production chains. 
The Markov chains are thinned by keeping every 15th sample and then flattened.

\subsubsection{Inferred posterior distributions}
\label{sec:bayes_results}

Next, we perform Bayesian parameter estimation as outlined in the previous section using the emulator and the FOM solver separately, and then compare the results to assess the emulator's efficacy.
Since we found in Sec.~\ref{sec:emu_scatt_obs} that the G-ROM and LSPG-ROM behave very similarly in emulating total cross sections, we focus here on the G-ROM. 
The corresponding results for the LSPG-ROM are available in our GitHub repository~\cite{BUQEYEsoftware}.
As in Sec.~\ref{sec:emu_scatt_obs}, we train the G-ROM using the greedy algorithm initialized with two random snapshots and a pool of 500 random candidate snapshots to scan the nine-dimensional LEC space during the greedy iteration.
We then let the greedy algorithm enrich the emulator basis until the requested error tolerance of $\tol = 10^{-2}$ is met, which must be done only once.
Note that we have chosen a large tolerance $\alpha$ to explore the case in which the EFT truncation error~\eqref{eq:eft_err_sigma} does not dominate the total theoretical uncertainty. 
This simulates a situation such as three-body scattering~\cite{Gnech:2025gsy,Gnech:2025lbg} where one can afford only relatively large emulator errors because otherwise the emulator training may be prohibitively slow.

\begin{figure}
    \centering
    \includegraphics[width=\linewidth]{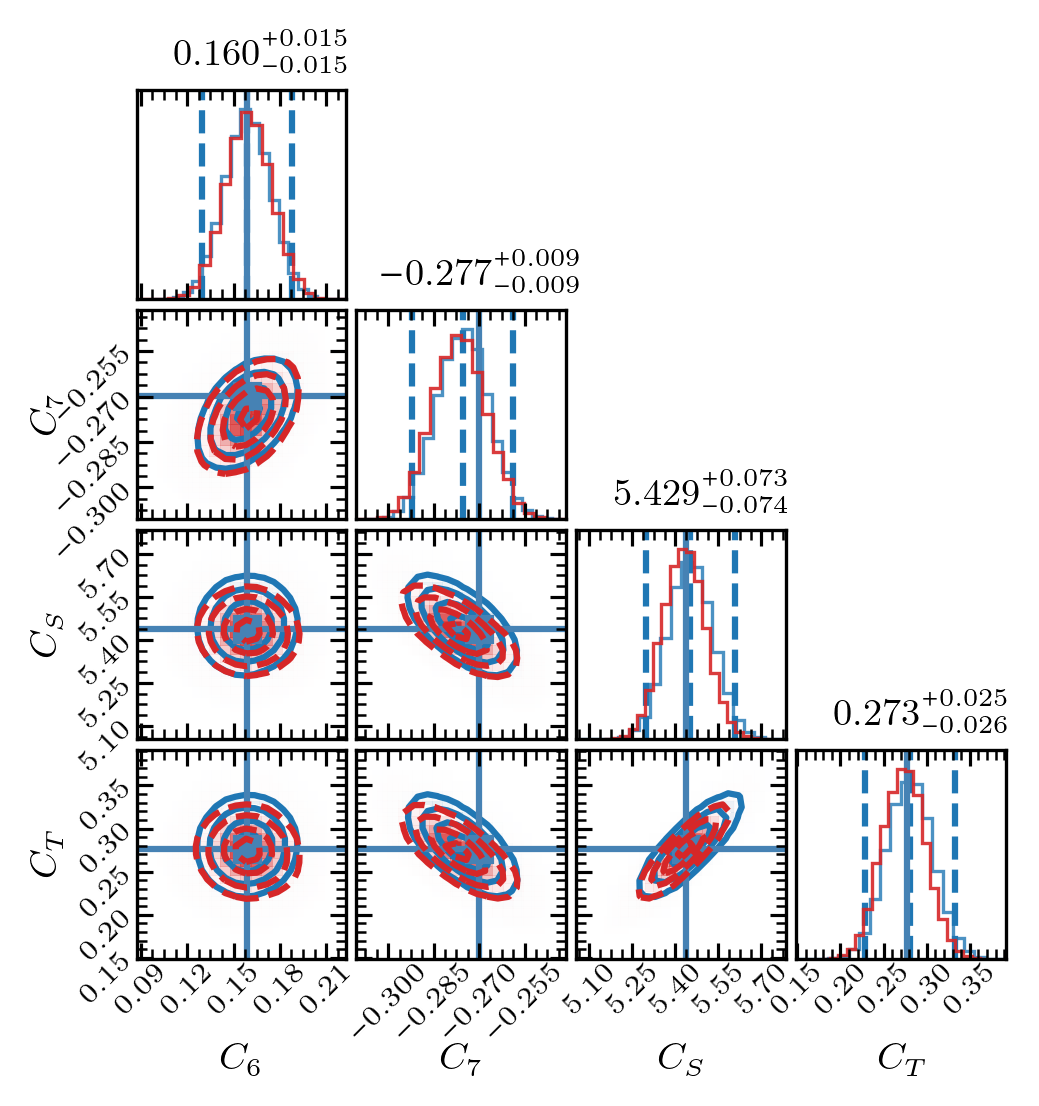}
    \caption{%
    \rev{Subspace of the inferred posterior distribution~\eqref{eq:posterior} of the LECs in the GT+ potential optimized to the PWA93 total cross section~\cite{Stoks:1993tb}. 
    The results based on the emulator (FOM solver) are shown in red (blue), with the realistic $\bar{c} = 1$ in Eq.~\eqref{eq:eft_err_sigma}.
    For brevity, we show here only a representative subspace of the nine-dimensional posterior distribution~\eqref{eq:posterior}, consisting of the LECs $(C_S, C_T, C_6, C_7)$.
    The blue lines and squares denote the best-fit values obtained in Ref.~\cite{Gezerlis:2014zia}.
    The titles of the marginal distributions (along the diagonal) and the dashed blue lines indicate the 95\% credibility interval centered on the median.%
    }%
    }
    \label{fig:posterior_lecs_full}
\end{figure}

For this specific case, Fig.~\ref{fig:posterior_lecs_full} shows the results of our MC sampling of the LEC posterior distribution~\eqref{eq:posterior} based on the emulator (red lines) and high-fidelity solver (blue lines).
All nine LECs of the GT+ potential are calibrated to total cross section data~\cite{Stoks:1993tb} at twelve laboratory energies, eleven evenly spaced in $E_\text{lab} = 0.1 \ldots 25.1 \MeV$ in increments of $2.5 \MeV$ (where the cross section changes rapidly with the energy) and an additional one at $60 \MeV$.
All of these points are collected in $\mathcal{D}$.

Specifically, Fig.~\ref{fig:posterior_lecs_full} shows the correlation plot of a three-dimensional subspace of the inferred nine-dimensional LEC posterior~\eqref{eq:posterior}.
For brevity, we present only the representative subspace comprising $(C_S, C_T, C_6, C_7)$.
The interested reader may find the complete correlation plot in the accompanying GitHub repository~\cite{BUQEYEsoftware}.
The blue horizontal and vertical lines, as well as the associated squares, denote the best-fit values obtained in Ref.~\cite{Gezerlis:2014zia} using a different fit protocol, e.g., by fitting to scattering phase shifts, not the total cross section as done here.
The titles of the marginal distributions (in the diagonal panels) report 95\% credibility intervals (dashed blue lines) centered on each distribution's median. 
For each random sample of the LEC posterior, we emulate the total cross section and check that the unitary violation of the $S_\ell$-matrix in all contributing partial-wave channels is $|1 - \eta_\ell| \lesssim 10^{-2}$. 
As shown in Fig.~\ref{fig:posterior_lecs_full}, even with the purposely large chosen emulator error (compared to the EFT truncation error) the LEC inference is not significantly distorted.

\subsubsection{Computational efficiency of the emulators}
\label{sec:emulator_speedups}
\begin{figure}[tb]
    \centering
    \includegraphics[width=\columnwidth]{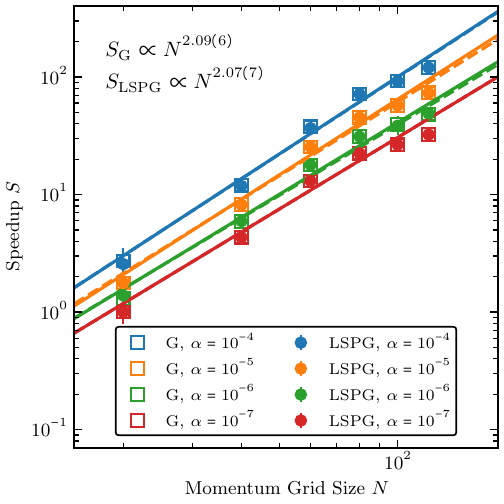}
    \caption{%
    \rev{Computational efficiency of the G-ROM (``G'') and LSPG-ROM (``LSPG'') in emulating the total cross section~\eqref{eq:tot_cs} based on partial waves with $j \leqslant 6$, shown as a function of the number of quadrature grid points used in the LS equation. 
    All timings are carried out over four energies simultaneously: 
    $E_{\text{lab}} = 10, 50, 100,$ and $200 \MeV$. 
    speedup factors for the two ROMs and four different requested tolerances $\tol$ for the emulated $\vb{t}$ are shown (see also the legend).
    The annotations report the results of least-squares fits to the data on a log-log scale to extract the scaling exponent. 
    For a typical grid size of $N=100$ (dashed vertical line), both achieve speedup factors ranging from $S\approx 26$ to $S\approx 72$, depending on the tolerance.
    For the largest grid size of $N=120$, and the largest tolerance $\tol = 10^{-4}$, the speedup increases further to $S \approx 92$. %
    }%
    }
    \label{fig:emulator_speedups}
\end{figure}
\begin{figure*}[tb]
    \centering
    \includegraphics[width=\textwidth]{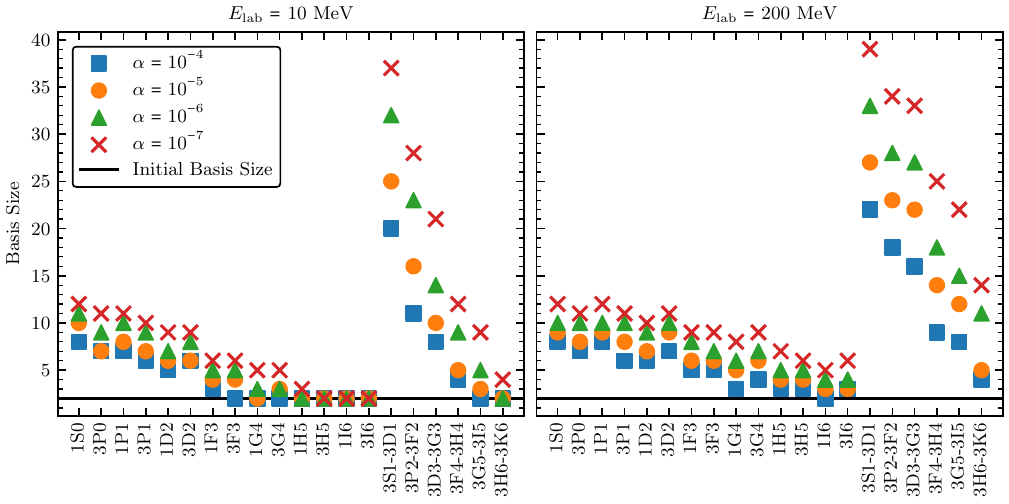}
    \caption{%
    Basis sizes produced by the G-ROM emulator for each partial-wave channel at $E_{\text{lab}} = 10 \MeV$ (left panel) and $E_{\text{lab}} = 200 \MeV$ (right panel), shown for several requested tolerance values (see the legend) and a fixed number of momentum grid points $N=100$. 
    At low energy, the required basis size decreases rapidly with increasing $j$, consistent with the suppression of high partial waves. 
    At $E_{\text{lab}} = 200 \MeV$, this decay is much weaker, and the high-$j$ channels require larger bases. 
    As expected, stricter tolerances $\tol$ lead to larger basis sizes for all channels. 
    The LSPG-ROM produces a similar pattern of basis sizes across channels, tolerances, and energies, closely mirroring the G-ROM results. %
    }
    \label{fig:basis_size}
\end{figure*}

\begin{figure}[tb]
    \centering
    \includegraphics[width=\columnwidth]{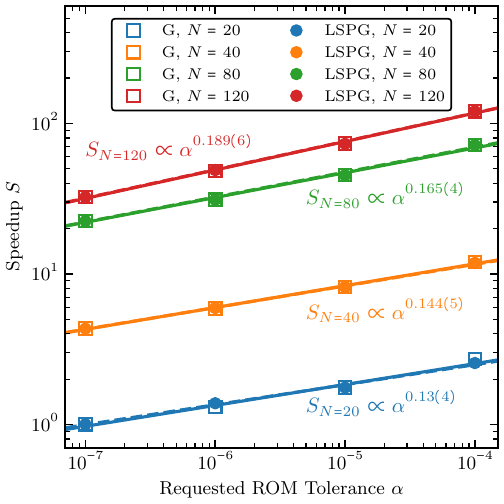}
    \caption{%
    The G-ROM and LSPG-ROM speedup factor~\eqref{eq:speedup_factor} follows a power-law dependence on the requested tolerance $\tol$ for grid sizes $N=20, 40, 80$, and $120$, with larger grid sizes exhibiting slightly steeper exponents.
    Consequently, decreasing the tolerance by an order of magnitude reduces the speedup by roughly $26-35\%$. 
    See Fig.~\ref{fig:emulator_speedups} and the main text for details.%
    }
    \label{fig:emulator_speedups_tol}
\end{figure}

Having established the accuracy of the developed emulators, we now assess their computational efficiency. 
The efficiency of each ROM relative to the FOM is quantified by the speedup factor, 
\begin{equation} \label{eq:speedup_factor}
    S = \frac{T_\text{FOM}}{T_\text{ROM}} \,,
\end{equation}
where $T_\text{FOM}$ and $T_\text{ROM}$ denote the average CPU runtimes per evaluation of the total cross section, Eq.~\eqref{eq:tot_cs}, at the best-fit LEC values. 
To obtain representative timings, we evaluate the total cross section at $E_\text{lab} = 10, 50, 100,$ and $ 200 \MeV$, and average the total runtime over 100 repetitions. 
All uncertainties are reported as standard errors of the mean across these repetitions. 
All partial-wave channels with $j \leqslant 6$ are included in the calculations. 
By definition, $S > 1$ indicates that the ROM is computationally more efficient than the FOM for the same task.

To obtain a clearer picture of the emulator performance and to provide an HPC-ready implementation suitable for Bayesian workflows, we implement both ROMs and the LS-based FOM solver in Python using \jax~\cite{jax2018github}. 
Leveraging an efficient offline-online decomposition together with just-in-time compilation and automatic vectorization offered by \jax, all potential matrix elements for all energies and partial-wave channels are precomputed and stored in the offline stage, as described in Sec.~\ref{sec:chiral_pot}. 
This procedure allows the FOM to evaluate the total cross section at many energies and channels in a single vectorized operation, providing an optimized baseline for comparing ROM speedups. 
Because the trained ROMs have different basis sizes across energy and partial-wave channels, we combine all individual $t$-matrix emulators into a unified array structure, enabling simultaneous evaluation in vectorized form. 
Finally, each ROM evaluation includes an on-the-fly estimate of the emulator error, which approximately doubles the per-evaluation run time; the reported speedup factors therefore account for this additional cost. 

Similar to Refs.~\cite{Garcia:2023slj,Maldonado:2025ftg}, we emphasize that absolute speedup factors are hardware- and implementation-dependent, and should therefore be interpreted with care.
To mitigate this sensitivity, Fig.~\ref{fig:emulator_speedups} examines the trend of the speedup factor $S(N)$ as a function of the number $N$ of quadrature points.
This scaling trend mainly reflects how much extra work each method requires when the problem becomes larger, which is a property of the method itself. 
As a result, the scaling curves are less affected by processor speed, memory layout, or compilation strategy, and the comparison becomes more general and easier to interpret. 

Figure~\ref{fig:emulator_speedups} displays the resulting $S(N)$ for the G-ROM (``G'') and LSPG-ROM (``LSPG'') across four requested tolerances for the emulated half-on-shell $t$-matrix vector $\vb{t}$ (see the legend).  
For each case, we fit $S(N)\propto N^p$ using a linear function on a log-log scale to extract the scaling exponent $p$. 
The resulting exponents are subsequently averaged over all tolerances for each ROM type, and the approximate averaged scalings $S(N)$ are annotated accordingly.
Because the two ROMs produce nearly identical basis sizes at a given tolerance, both exhibit similar scaling behavior, $p \approx 2$.
That is, $S(N)$ approximately quadruples if $N$ is doubled.
For the largest grid size $N=120$ shown in Fig.~\ref{fig:emulator_speedups}, we find $S(N) \approx \mathcal{O}(10^3)$ at the highest tolerance of $\tol = 10^{-4}$, indicating the ROMs' overall high computational efficiency

To better understand the origin of these trends, Fig.~\ref{fig:basis_size} shows the number of greedy-selected snapshots required by the G-ROM at $E_{\text{lab}} = 10$ (left panel) and $200 \MeV$ (right panel) for the same tolerance values $\tol$ as in Fig.~\ref{fig:emulator_speedups}, providing a direct view of how the complexity of each channel grows with energy and requested accuracy. 
At low energies, the decrease in basis size with increasing $j$ reflects the suppression of high partial waves, resulting in compact reduced bases for most channels. 
As the energy increases, this suppression weakens and more basis vectors are required to accurately represent the scattering dynamics, particularly in the highest partial-wave channels considered. 
For coupled channels, the snapshot vectors are four times longer than those of single channels, leading to reduced bases that are roughly four times larger at fixed $j$.

These basis-size trends play a central role in determining the overall computational cost of the ROM evaluations.
As expected, decreasing the requested tolerance $\tol$ systematically increases the number of snapshots for all channels, reflecting the tighter accuracy requirements imposed on the reduced representation. Additionally, because the reduced basis size depends strongly on the laboratory energy and the ROM evaluations are vectorized over channels and energies, the overall speedup factor is typically limited by the coupled channels at the highest energy, which set the dominant computational bottleneck. 

Figure~\ref{fig:emulator_speedups_tol} shows the emulator speedup factor $S(\tol)$ as a function of the requested ROM tolerance $\tol$ for the momentum grid sizes $N=20, 40, 80$, and $120$. 
Across all $N$, the curves are nearly linear on a log-log scale, indicating a power-law dependence of the form $ S(\tol) \propto \tol^p$, with $p$ ranging from 0.13(4) for $N=20$ up to 0.189(6) for $N=120$. 
This relationship means that tightening the tolerance by one order of magnitude reduces the speedup by a fixed multiplicative factor $10^{-p}$, corresponding to a fractional decrease of $1-10^{-p}$. 
The larger $N$ grids exhibit slightly larger exponents, and therefore a more pronounced decrease in speedup as the tolerance $\tol$ becomes stricter. 
For example, for $N=20$, lowering the tolerance by a factor of 10 reduces the speedup by about $1-10^{-0.13} \approx 0.26$, or 26\%. 
For $N=120$, the reduction is closer to 35\%. 
Thus, while the grid must be sufficiently large to realize substantial speedups, the cost of tightening the tolerance also grows mildly with $N$. 
Even so, the emulator ultimately delivers its largest overall speedups at the largest grid sizes. 

\section{Summary and Outlook}
\label{sec:summary}

In this work, we have extended the active learning emulators G-ROM and LSPG-ROM, developed by Maldonado et al.~\cite{Maldonado:2025ftg} for two-body scattering with error estimation in coordinate space, to coupled-channel scattering in momentum space.
\rev{Our \rev{FOM} solver is based on the Lippmann-Schwinger integral equation for the scattering $t$-matrix as opposed to the radial Schr{\"o}dinger equation.}
Hence, our approach gives access to a much wider range of modern chiral NN interactions, including the high-order SMS potentials developed by LENPIC~\cite{Reinert:2017usi,Epelbaum:2022cyo}.

Specifically, our emulators are based on Galerkin (G-ROM) and Petrov-Galerkin (LSPG-ROM) projections, as well as on high-fidelity calculations at a few snapshots across the parameter space of the chiral NN interactions~\cite{Melendez:2022kid,Drischler:2022ipa}. 
Chiral NN interactions exhibit an affine dependence on the short-range contact interactions, which we considered here as parameters, enabling efficient offline-online decompositions.
A greedy algorithm estimates the emulator error and iteratively enriches the emulator basis by adding snapshots to regions where the estimated error is largest, until a specified error tolerance is met~\cite{hesthaven2015certified,Quarteroni:218966,Benner20201}.
Similarly to Ref.~\cite{Maldonado:2025ftg}, we found that this active learning approach to emulator training has two significant advantages over the na{\"i}ve POD approach: 
It typically requires much fewer (computationally demanding) high-fidelity calculations to achieve similar accuracy, and includes emulator error estimation.
Both features are critical for conducting principled uncertainty quantification of scattering.

Overall, we have found that the greedy algorithm performs similarly well in momentum space for constructing ROMs as in coordinate space~\cite{Maldonado:2025ftg}.
For example, it converges exponentially when applied to the GT+ potential in the uncoupled \oneSzero and the coupled \threeSDone channels, with accuracies comparable to those of the POD approach.
After this convergence analysis, we presented results for emulating phase shifts in the deuteron channel (as a representative channel) and total cross sections.
The latter calculations involve emulating phase shifts across various (coupled and uncoupled) partial-wave channels.
In all cases, the emulators reproduced the high-fidelity calculations with high accuracy.

We propagated the emulator error (in the scattering $t$-matrix vector) to the total cross section prediction in addition to the estimated EFT truncation error.
We then demonstrated, in a proof-of-principle calculation, the efficacy of these active learning emulators with error estimation for Bayesian parameter estimation of the LECs for a chiral interaction, using total cross-section data.
In the future, this approach allows one to treat the emulator error on the same footing as other theoretical errors, such as the EFT truncation error.

Emulator speedup factors depend inherently on the software and hardware implementations. 
To assess the computational speedup factors more rigorously, without being overly exhaustive, we implemented highly efficient FOMs and ROMs in Python using \jax.
We have found that our emulators can be $100$ times faster (or even more) than the already efficient FOM solver, depending on the emulator's basis size, enabling efficient Bayesian parameter estimation, even at the two-body level.

Our emulators, together with our recent advances in developing active learning emulators for 3N scattering~\cite{Gnech:2025gsy, Gnech:2025lbg}, constitute a significant step toward comprehensive Bayesian calibrations of chiral NN and 3N forces using scattering data.
These Bayesian analyses will be critical for elucidating issues in chiral EFT~\cite{Furnstahl:2021rfk}, including the efficacy of modified power countings relative to the popular Weinberg power counting and the differing predictions of chiral EFT for ground-state energies and charge radii of atomic nuclei in the medium- to heavy-mass regime. 

Future work will focus on combining the developed NN and 3N emulators into a unified framework for systematic Bayesian analysis of scattering observables. 
To achieve this goal, several developments are necessary.
Our emulators with error estimation need to be extended to other scattering observables, such as differential cross sections and spin observables.
Furthermore, to emulate non-affine model parameters, such as the momentum cutoff and pion mass, versatile hyperreduction methods, including rational function approximations, need to be developed to enable efficient emulator offline-online decompositions.
At the NN level, the empirical interpolation method (EIM) has already been tested and demonstrated to be efficient at rendering Woods-Saxon-like potentials approximately affine~\cite{Odell:2023cun,Catacora-Rios:2025pau}.

In addition, accurate and efficient methods for estimating the minimum singular values of the (asymmetric) FOM matrix $\vb{A}(\paramVec)$ in the parameter space need to be developed to derive more rigorous upper bounds on the emulator errors
(see also Ref.~\cite{Maldonado:2025ftg}).
These would enable conservative error estimation.
Furthermore, the 3N emulators developed in Refs.~\cite{Gnech:2025gsy,Gnech:2025lbg} need to be extended to energies above the deuteron breakup threshold, a project that is currently under development by the STREAMLINE2 collaboration.

The ultimate goal is to apply this unified framework for emulating scattering observables to calibrate next-generation chiral NN and 3N forces, with emulator errors and EFT truncation errors fully quantified using the BUQEYE machinery~\cite{Melendez:2017phj,Melendez:2019izc,Wesolowski:2021cni,Svensson:2022kkj,Svensson:2023twt}, and study their predictions for atomic nuclei and infinite matter.
These developments will benefit from the software framework we \rev{made publicly available~\cite{BUQEYEsoftware}}, thereby enabling broader adoption of active learning emulators with error estimation within the nuclear physics community. 

\begin{acknowledgments}

We thank Petar Mlinari{\'c}, Stephan Rave, and our STREAMLINE2 collaborators for fruitful discussions, and Ingo Tews for sharing the source code that evaluates the chiral potentials developed in Ref.~\cite{Gezerlis:2014zia}.
We also thank ECT* for support at the 2025 workshop ``Next generation ab initio nuclear theory'' during which this work was presented and further developed.
This material is based upon work supported by the U.S. Department of Energy, Office of Science, Office of Nuclear Physics under contract No.~DE-FG02-93ER40756, under the FRIB Theory Alliance award DE-SC0013617, under the STREAMLINE collaboration awards DE-SC0024233 (Ohio University) and DE-SC0024509 (Ohio State University), the STREAMLINE2 collaboration award DE-SC0026198, and by the National Science Foundation under awards PHY-2209442/PHY-2514765.
Computational resources for this work were provided by the NSF MRI grant 2320493 and the National Energy Research Scientific Computer Center (NERSC), a U.S. Department of Energy Office of Science User Facility supported by the Office of Science of the U.S. Department of Energy under Contract No.~DE-AC02-05CH11231 using NERSC award NP-ERCAP0033451.
The following open-source libraries were used to generate the results in this work:
\texttt{corner}~\cite{corner},
\texttt{cython}~\cite{behnel2011cython},
\texttt{emcee}~\cite{Foreman_Mackey:2013aa},
\jax~\cite{jax2018github},
\texttt{GSL}~\cite{gsl},
\texttt{matplotlib}~\cite{Hunter:2007},
\texttt{mpmath}~\cite{mpmath},
\texttt{numpy}~\cite{harris2020array}, and
\texttt{scipy}~\cite{2020SciPy-NMeth}.

\end{acknowledgments}

\appendix

\section{Explicit calculation and implementation of the LS equation for NN scattering}
\label{appendixA}

In this appendix, we present the explicit numerical solution of the single and coupled Lippmann-Schwinger (LS) integral equations
and write them in the algebraic form that is most convenient for setting up the emulators.

\subsection{Single channel equation}
Using the nonrelativistic expression for the energies, $E=k^2/2\mu$, Eq.~\eqref{eq:2.2} reads
\begin{align}
&T^j_\ell(k, k';E) = V^j_\ell(k, k')  \notag \\
 & \qquad \null +
 \lim_{\varepsilon \to 0}  \int_0^\infty dk'' \, k''^2 \frac{2\mu \, V^j_\ell(k, k'') T^j_\ell(k'', k'; E)}{k_0^2 - k''^2 + i\varepsilon}. 
\label{eq:A.1}
\end{align}
Using the Sokhotski-Plemelj theorem, 
\begin{equation}
\lim_{\varepsilon \to 0} \frac{1}{x_0 - x + i\varepsilon} = \pv{\frac{1}{x_0 - x}} - i \pi \delta(x_0 - x),
\end{equation}
we obtain
\begin{align}
T^j_\ell(k, k'; k_0) &= V^j_\ell(k,k')
+ \pv{\int_0^\infty} dk'' \frac{f(k'')}{k_0^2 - k''^2}
\notag \\
& \qquad \null
 - i \pi \int_0^\infty dk'' \delta(k_0^2 - k''^2) f(k''), 
\label{eq:A.2}
\end{align}
where we define a function $f(k'')$ as
\begin{equation}
f(k'') = 2\mu \, k''^2 \, V^j_\ell(k, k'') T^j_\ell(k'', k'; k_0).
\end{equation}
This simplifies Eq.~(\ref{eq:A.2}) to
\begin{align}
T_\ell(k, k', k_0) &=V_\ell(k, k') + \pv{\int_0^{q_\text{max}}} dk'' \frac{f(k'')}{k_0^2 - k''^2} 
 \notag \\
 & \qquad \null - i \pi \frac{f(k_0)}{2 k_0}. 
\label{eq:A.5}
\end{align}
For convenience, we will from now on omit the superscript $j$ indicating the total angular momentum channel.

For computational purposes, the upper limit of the integral in Eq.~\eqref{eq:A.5} is truncated at a fixed $q_\text{max}$.
The principal value integral can now be split as,
\begin{align}
T_\ell(k, k', k_0) &= V_\ell(k, k') + \int_0^{q_\text{max}} dk'' \frac{f(k'') - f(k_0)}{k_0^2 - k''^2} \nonumber \\
&\quad + \frac{f(k_0)}{2 k_0} \ln \frac{q_\text{max} + k_0}{q_\text{max} - k_0} - i \pi \frac{f(k_0)}{2 k_0}.
\end{align}
The singularity at \( k'' = k_0 \) is removed by subtracting its value in the first integral, and then adding it back in. The added term 
 \( f(k_0) \)  can be evaluated analytically, leading to
\begin{equation}
\pv{\int_0^{q_\text{max}}} dk'' \frac{f(k_0)}{k_0^2 - k''^2} = \frac{f(k_0)}{2 k_0} \ln \frac{q_\text{max} + k_0}{q_\text{max} - k_0}.
\end{equation}
It should be noted that when $q_\text{max} \rightarrow \infty$, the logarithmic term vanishes.
The value of $q_{\rm max}$ needs to be chosen such that the evaluation of the integral is independent of it. For chiral interactions, this choice can be informed by the momentum cutoff applied in the interaction.

To further proceed, we discretize the LS equation on a grid of $N$ points determined by an integration grid of our choice $k \rightarrow i $, $k' \rightarrow j$, and $k'' \rightarrow i'$.
Using the above notation, the discretized LS equation reads,
\begin{align}
 V_\ell^{ij} &= \int dk_{i'} \delta(k_{i'} - k_i) T_\ell^{i'j} 
 -  \biggl[ \int_0^{q_\text{max}} dk_{i'} \frac{f(k_{i'}) - f(k_0)}{k_0^2 - k_{i'}^2} 
 \notag \\
 & \qquad \null
  + \Bigl( \ln\frac{q_\text{max} + k_0}{q_\text{max} - k_0} - i \pi \Bigr) \frac{f(k_0)}{2 k_0} \biggr] 
  \notag \\
&=  \sum_{i' =1}^n \Bigl(\delta_{i'i} - \frac{2 \mu w_i'}{k_0^2 - k_{i'}^2} k_{i'}^2 V_\ell^{ii'} \Bigr)T_\ell^{i'j} \notag \\
& \null \qquad -  \Bigl[ -k_0^2 \sum_{i'=1}^n \frac{2 \mu w_i'}{k_0^2 - k_{i'}^2} + \mu k_0 \Bigl( \ln\frac{q_\text{max} + k_0}{q_\text{max} - k_0} - i \pi \Bigr) \Bigr]  \notag \\
& \qquad \null \times  V_\ell^{i',n+1}T_\ell^{n+1,j},
\label{eq:A.8}
\end{align}
with
\begin{equation}
        g_i = \frac{2 \mu w_i}{k_0^2 - k_i^2},
        \label{eq:A.9}
\end{equation}
and
\begin{equation}
        b := -k_0^2 \sum_{i=1}^n g_i + \mu k_0 \left( \ln\frac{q_\text{max} + k_0}{q_\text{max} - k_0} - i \pi \right),
        \label{eq:A.10}
\end{equation}
In Eq.~(\ref{eq:A.9}), the quantities $w_i$ represent the integration weights.
Equation~(\ref{eq:A.8}) can now be written in short form as
\begin{eqnarray}
        V_\ell^{ij} &=& \sum_{i'=1}^n \left( \delta_{i'i} - g_{i'} k_{i'}^2 V_\ell^{ii'} \right) T_\ell^{i'j} + \left(  - b V_\ell^{i,n+1} \right) T_\ell^{n+1,j} \cr
 &=& \sum_{i'=1}^n A_{ii'} T_\ell^{i'j} + A_{i,n+1} T_\ell^{n+1,j} \cr
 &=& \sum_{i'=1}^{n+1} A_{ii'} T_\ell^{i'j},
 \label{eq:A.11}
\end{eqnarray}
where
\begin{equation}
  A_{ii'} \equiv
  \begin{cases}
  \delta_{ii'} - g_{i'} k_{i'}^2 V_\ell^{ii'},    & 1 \leq i' \leq n, \\
                
  - b V_\ell^{i,n+1},   & j = n+1.
  \end{cases}
\end{equation}
With these manipulations, the single-channel LS equation becomes a matrix equation of the form
\begin{equation}
    A \, T = V,
    \label{eq:A.12}
\end{equation}
which can be solved by standard linear algebra routines. For the high-fidelity solutions, we use the 
\texttt{SciPy}'s wrapper around LAPACK's efficient solver \texttt{ZGESV} for general matrices, \texttt{linalg.solve()}~\cite{2020SciPy-NMeth}.

\subsection{Coupled channel equations}

The LS equations for the coupled channels are defined in Eq.~(\ref{eq:2.3}). Similar to the single-channel equation, they can be written as
\begin{widetext}
\begin{align}
        &V_{\ell \ell'}^{ij} = \int dk_{i'} \delta(k_{i'} - k_i) T_{\ell'' \ell'}^{i'j} - \sum_{\ell''}\left[ \int_0^{q_\text{max}} dk_{i'} \frac{f(k_{i'}) - f(k_0)}{k_0^2 - k_{i'}^2}
                + \left( \ln\frac{q_\text{max} + k_0}{q_\text{max} - k_0} - i \pi \right) \frac{f(k_0)}{2 k_0} \right] \cr
  &= \sum_{\ell''}\sum_{i' =1}^n \left(\delta_{i'i}\delta_{\ell'' \ell} - \frac{2 \mu w_i'}{k_0^2 - k_{i'}^2} k_{i'}^2 V_{\ell \ell''}^{ii'} \right)T_{\ell'' \ell'}^{i'j} - \sum_{\ell''}   \left[ -k_0^2 \sum_{i'=1}^n \frac{2 \mu w_i'}{k_0^2 - k_{i'}^2} + \mu k_0 \left( \ln\frac{q_\text{max} + k_0}{q_\text{max} - k_0} - i \pi \right) \right] V_{\ell \ell''}^{i',n+1}  T_{\ell'' \ell'}^{n+1,j}, \nonumber
  \label{eq:A.14}
\end{align}
\end{widetext}
where
\begin{equation}
A_{\ell \ell''}^{ii'} \equiv
\begin{cases}
\delta_{ii'} \delta_{\ell'' \ell} - g_{i'} k_{i'}^2 V_{\ell \ell''}^{ii'}, & 1 \leq i' \leq n, \\
- b V_{\ell \ell''}^{i,n+1}, & k = n+1\,.
\end{cases}
\end{equation}
\\

With $S=1$, we have the different combinations of $\ell, \ell' = \{{j-1 := m}, {j+1 := p}$\} that lead to four different situations: \\
Case $\ell = m, \ell' = m$:
\begin{align}
    &V_{mm}^{ij} = \sum_{i'=1}^{n+1} A_{mm}^{ii'} T_{mm}^{i'j} + \sum_{i'=1}^{n+1} A_{mp}^{ii'} T_{pm}^{i'j} \,,\cr
    &V_{mm} =  A_{mm} T_{mm}+ A_{mp} T_{pm}\,.
\end{align}
Case: $\ell = p, \ell' = m$:
\begin{align}
    &V_{pm}^{ij} = \sum_{i'=1}^{n+1} A_{pm}^{ii'} T_{mm}^{i'j} + \sum_{i'=1}^{n+1} A_{pp}^{ii'} T_{pm}^{i'j} \,,\cr
    &V_{pm} = A_{pm} T_{mm} +  A_{pp}T_{pm}\,.
\end{align}
Case $\ell = m, \ell' = p$:
\begin{align}
    &V_{mp}^{ij} = \sum_{i'=1}^{n+1} A_{mm}^{ii'} T_{mp}^{i'j} + \sum_{i'=1}^{n+1} A_{mp}^{ii'} T_{pp}^{i'j} \,,\cr
    &V_{mp} = A_{mm} T_{mp} +  A_{mp}T_{mm}\,.
\end{align}
Case $\ell = p, \ell' = p$:
\begin{align}
    &V_{pp}^{ij} = \sum_{i'=1}^{n+1} A_{pm}^{ii'} T_{mp}^{i'j} + \sum_{i'=1}^{n+1} A_{pp}^{ii'} T_{pp}^{i'j} \,,\cr
    &V_{pp} = A_{pm} T_{mp} +  A_{pp}T_{pp}\,.
\end{align}
Combining them leads to a $2(N+1)\times 2(N+1)$ dimensional matrix equation
\begin{equation}
\begin{pmatrix}
A_{mm} &  A_{mp} \\
A_{pm} & A_{pp}
\end{pmatrix} \,  \begin{pmatrix}
T_{mm} &  T_{mp} \\
T_{pm} & T_{pp}
\end{pmatrix} = \begin{pmatrix}
V_{mm} &  V_{mp} \\
V_{pm} & V_{pp},
\end{pmatrix}\,,
\label{eq:A.19}
\end{equation}
which again is a system of linear equations of the type $ A \, T = V$ to be solved using standard linear algebra packages (in our case, we use \texttt{ZGESV} from LAPACK).

For the purpose of constructing the emulator, we need to rewrite the coupled-channel $t$-matrix as a single vector of length $4(N+1)$. Thus, in the actual calculation, Eqs.~(\ref{eq:A.19}) take the form
\begin{equation}
\begin{pmatrix}
A_{mm} & A_{mp} &   \vb{0} & \vb{0} \\
A_{pm} & A_{pp} &   \vb{0} & \vb{0} \\
\vb{0} &  \vb{0} & A_{mm}   &  A_{mp} \\
\vb{0} &  \vb{0} & A_{pm}   &  A_{pp} \\
\end{pmatrix} \,  \begin{pmatrix}
T_{mm}\\
T_{pm} \\
T_{mp} \\
T_{pp}
\end{pmatrix} = \begin{pmatrix}
V_{mm} \\
V_{pm} \\
V_{mp}\\
V_{pp}
\end{pmatrix}.
\label{eq:A.20}
\end{equation}
Here, each of the matrices $A$ has the dimension $(N+1)\times (N+1)$, and $\vb{0}$ stands for the zero matrix of the same dimension.

\section{Calculation of phase shifts, inelasticity parameters, and total cross sections}
\label{appendixB}

In this appendix, we present the calculation of phase-shifts,  inelasticity parameters, and total cross sections from the on-shell $t$-matrix elements. For completeness, we also present the explicit diagonalization of the coupled-channels according to the Stapp~\cite{Stapp:1956mz} parameterization.  We conclude with the expression for the total cross section.

To calculate the two-nucleon phase shifts and inelasticity parameters, we start from the expression for the partial wave $S$-matrix elements given in Eq.~(\ref{eq:2.5}) and first consider single-channel scattering.
For $\ell =\ell'$, i.e. the singlet and triplet channels, the scattering amplitude $\tau^j_{\ell \ell'}(E_{k_0})$ can be written as
\begin{equation}
\tau^j(E_{k_0}) = \frac{\eta^j(E_{k_0}) e^{2i\delta^j(E_{k_0})}-1}{2i},
\label{eq:B.1}
\end{equation}
where the indices $\ell, \ell'$ are suppressed for convenience. For the real and imaginary parts, one obtains
\begin{eqnarray}
\Re  \tau^j(E_{k_0})& =& \frac{\eta^j(E_{k_0})}{2} \sin 2\delta^j(E_{k_0}) , \cr
\Im  \tau^j(E_{k_0})& =& \frac{1}{2} - \frac{\eta^j(E_{k_0})}{2} \cos 2\delta^j(E_{k_0}) .
\label{eq:B.2}
\end{eqnarray} 
As an aside, plotting $\Re  \tau^j(E_{k_0})$ versus $\Im  \tau^j(E_{k_0})$ leads to the well-known Argand diagrams.
The phase shifts and inelasticities for the spin singlet and uncoupled spin triplet states are directly obtained from 
Eqs.~(\ref{eq:B.2}) as
\begin{eqnarray}
\delta^j &=& \frac{1}{2} \arctan \frac{\Re  \tau^j}{\frac{1}{2} -\Im  \tau^j} , \\
\eta^j &=& \sqrt{ 1+4\left[ (\Re \tau^j)^2 + (\Im \tau^j)^2 -\Im \tau^j)\right] }. 
\label{eq:B.3}
\end{eqnarray}
For clarity of notation, we also omitted the dependence of the phase shift and the inelasticity on the energy.

For the coupled channel states, the partial wave $S$-matrix elements can be parameterized in the Stapp~et~al.~\cite{Stapp:1956mz} form, using the ``bar'' phase shifts $\bar{\delta}_-$, $\bar{\delta}_+$ and the mixing parameter $\bar{\varepsilon}$ by means of the transformation
\begin{eqnarray}
\lefteqn{
\begin{pmatrix}
S^j_{--} &  S^j_{+-} \\
S^j_{-+} & S^j_{++} \end{pmatrix} = } &&   \\
& & \begin{pmatrix}
e^{i\beta^j_-} &  0 \\
0 & e^{i\beta^j_+}
\end{pmatrix}  \begin{pmatrix}
\cos(2\Bar{\varepsilon}^j) &  i\sin(2\Bar{\varepsilon}^j) \\
i\sin(2\Bar{\varepsilon}^j) & \cos(2\Bar{\varepsilon}^j)
\end{pmatrix} \, \begin{pmatrix}
e^{i\beta^j_-} &  0 \\
0 & e^{i\beta^j_+}
\end{pmatrix}. \nonumber 
\label{eq:B4}
\end{eqnarray}
Here the subscripts $-$ and $+$ label the two coupled partial-wave channels with orbital angular momenta $\ell_- = j-1$ and $\ell_+ = j+1$, respectively, for a fixed total angular momentum $j$, and 
\begin{equation}
e^{i\beta^j_\pm} := \bar{\eta}^j_\pm e^{2i \bar{\delta}^j_\pm}.
\end{equation}
Evaluating Eq.~(\ref{eq:B4}) leads to 
\begin{eqnarray}
\Bar{\varepsilon}^j&=&\frac{1}{2} \arctan\left( \frac{-i  (S^j_{+-} + S^j_{-+})}{2  \sqrt{S^j_{--} S^j_{++}}}\right) , \cr
\bar{\delta}^j_\pm &=& \frac{1}{2} \arctan \frac{ \Im (S^j_{\pm \pm} / \cos (2\bar{\varepsilon}^j)}
                                  {\Re (S^j_{\pm \pm} / \cos (2\bar{\varepsilon}^j)}, \cr
\bar{\eta}^j_\pm &=& \left|\frac{S^j_{\pm \pm}} {\cos(2\Bar{\varepsilon}^j)}\right| .
\label{eq:B.6}
\end{eqnarray}
The calculations of the high-fidelity solutions in the present work have been validated by comparing them to the codes of Refs.~\cite{Elster:1988pp,Elster:1988zu}. 

Using the bar phase shifts and inelasticities, we can define amplitudes $\tau^j_\pm (E_{k_0})$ similar to Eq.~\eqref{eq:B.1}. 
We calculate the total cross sections using the optical theorem. In terms of scattering amplitudes $\tau^j$ it can be written as~\cite{Silbar:1979wp}
\begin{equation}
    \sigma_{\text{tot}} = -\frac{i\pi}{k_0^2} \sum_{j} (2j+1) \sum_k\tau^j_k (E_{k_0}), 
    \label{eq:B.7}
\end{equation}
where the sum with index $k$ runs over the singlet, triplet and $\pm$-channels.


\bibliographystyle{apsrev4-2}
\bibliography{bayesian_refs,more_refs,bib}

\end{document}